\documentclass[letterpaper,twocolumn,10pt]{article}

\usepackage{usenix}
\makeatletter
\def\myparagraph{\@startsection
     {paragraph}{4}{\z@}{.7ex}{-1em}{\normalsize\em}}
\def\paragraph{\@startsection
     {paragraph}{4}{\z@}{1ex}{-1em}{\normalsize\bf}}
\makeatother

\newcommand{\mytt}[1]{{\small\tt #1}}

\usepackage{xspace}
\usepackage{enumitem}
\usepackage[english]{babel}
\usepackage{graphicx}
\usepackage{adjustbox}
\usepackage{subfigure}
\usepackage{wrapfig}
\usepackage{lscape}
\usepackage{rotating}
\usepackage[disable]{todonotes}
\usepackage{multirow}
\usepackage{pbox}
\usepackage{listings}
\usepackage{xcolor}
\usepackage{amssymb}
 \usepackage{authblk}
\usepackage[colorlinks=true, linkcolor=blue, 
            citecolor=blue, urlcolor=blue,
            bookmarks=true,        %
            bookmarksnumbered=true,%
]{hyperref}

\usepackage{url}
\newcommand{\comment}[1]{}

\usepackage[font={footnotesize},labelfont=bf]{caption}
\usepackage{color}
\definecolor{editorGray}{rgb}{0.95, 0.95, 0.95}
\definecolor{editorOcher}{rgb}{1, 0.5, 0} %
\definecolor{editorGreen}{rgb}{0, 0.5, 0} %
\lstdefinelanguage{JavaScript2}{
  morekeywords={typeof, new, true, false, catch, try, function, return, null, catch, switch, var, if, in, while, do, else, case, break},
  morecomment=[s]{/*}{*/},
  morecomment=[l]//,
  morestring=[b]",
  morestring=[b]',
  moredelim=**[is][\bf]{@}{@}
 }

\newcommand{\hsp}{\hspace*{2em}}

\newcommand{\vtodo}[1]{\todo[inline]{Venkat: #1}} %
\newcommand{\vtodon}[1]{}

\newcommand{\sstodon}[1]{}

\newcommand{\stodon}[1]{}
\newcommand{\rtodo}[1]{\todo[inline]{Rigel: #1}}
\newcommand{\rtodon}[1]{}

\newcommand{\btodon}[1]{}

\newcommand{\projname}{{\sc Sleuth}\xspace}

\newcommand{\titlec}{\projname: Real-time~Attack~Scenario~Reconstruction from~COTS~Audit~Data} 

\usepackage{enumitem}
\setlist[itemize,1]{leftmargin=1.5\parindent, itemsep=0ex, topsep=0.5ex, %
}
\setlist[enumerate,1]{leftmargin=2\parindent, itemsep=0ex, topsep=0.5ex, %
}
\makeatletter
\newcommand\intermediate{\@setfontsize\intermediate\@xipt{13}}
\def\section{\@startsection {section}{1}{\z@}{-2ex}{.9ex}{\large \bf}}
\def\subsection{\@startsection{subsection}{2}{\z@}{-1.6ex}{1ex}{\intermediate\bf}}
\renewcommand*{\thetable}{\arabic{table}}
\renewcommand*{\thefigure}{\arabic{figure}}
\let\c@table\c@figure
\makeatother

\begin{document}
\title{\titlec\thanks{This work was primarily supported by  DARPA (contract FA8650-15-C-7561) and in part by NSF (CNS-1319137, CNS-1421893, CNS-1514472 and DGE-1069311) and ONR (N00014-15-1-2208 and N00014-15-1-2378).
The views, opinions, and/or findings expressed are those of the author(s) and should not be interpreted as representing the
 official views or policies of the Department of Defense or the U.S. Government.
}
}
\date{}
\author[1]{Md Nahid Hossain}
\author[2]{Sadegh M. Milajerdi}
\author[1]{Junao Wang}
\author[2]{Birhanu Eshete}
\author[2]{Rigel Gjomemo}
\author[1]{R.~Sekar}
\author[1]{Scott D. Stoller}
\author[2]{V.N. Venkatakrishnan}

\affil[1]{%
Stony Brook University}
\affil[2]{%
University of Illinois at Chicago}
\maketitle
\thispagestyle{empty}

\pagestyle{empty}

\begin{abstract}
We present an approach and system for real-time reconstruction of attack
scenarios on an enterprise host. To meet the scalability and real-time needs of
the problem, we develop a platform-neutral, main-memory based, dependency graph
abstraction of audit-log data. We then present efficient, tag-based
techniques for attack detection and reconstruction, including source identification and impact analysis.  We also develop methods to
reveal the big picture of attacks by construction of compact, visual graphs
of attack steps.
Our system participated in a red team evaluation
organized by DARPA and was able to successfully detect and
reconstruct the details of the red team's attacks on hosts running Windows, FreeBSD and Linux. 
\end{abstract}

\section{Introduction} \label{intro}
We are witnessing a rapid escalation in targeted cyber-attacks (``Enterprise
Advanced and Persistent Threats (APTs)'')~\cite{apt-reports} conducted by
skilled adversaries. By combining social engineering techniques (e.g.,
spear-phishing) with advanced exploit techniques, these adversaries routinely
bypass widely-deployed software protections such as ASLR, DEP and sandboxes. As
a result, enterprises have come to rely increasingly on second-line defenses,
e.g., intrusion detection systems (IDS), security information and event
management (SIEM) tools, identity and access management tools, and application
firewalls. While these tools are generally useful, they typically generate a
vast amount of information, making it difficult for a security analyst to
distinguish truly significant attacks --- the proverbial
``needle-in-a-haystack'' --- from background noise. Moreover, analysts lack the
tools to ``connect the dots,'' i.e., piece together fragments of an attack
campaign that span multiple applications or hosts and extend over a long time
period. Instead, significant manual effort and expertise are needed to piece
together numerous alarms emitted by multiple security tools. Consequently, many
attack campaigns are missed for weeks or even months
\cite{target,neiman-marcus}.

In order to effectively contain advanced attack campaigns, 
analysts need a new generation of tools that not only
assist with detection but also produce a compact summary of the causal chains
that summarize an attack. Such a summary would enable an analyst to quickly
ascertain whether there is a significant intrusion, understand how the
attacker initially breached security, and determine the impact of the attack.

\begin{figure*}[t]
\centering
    \includegraphics[width=\textwidth]{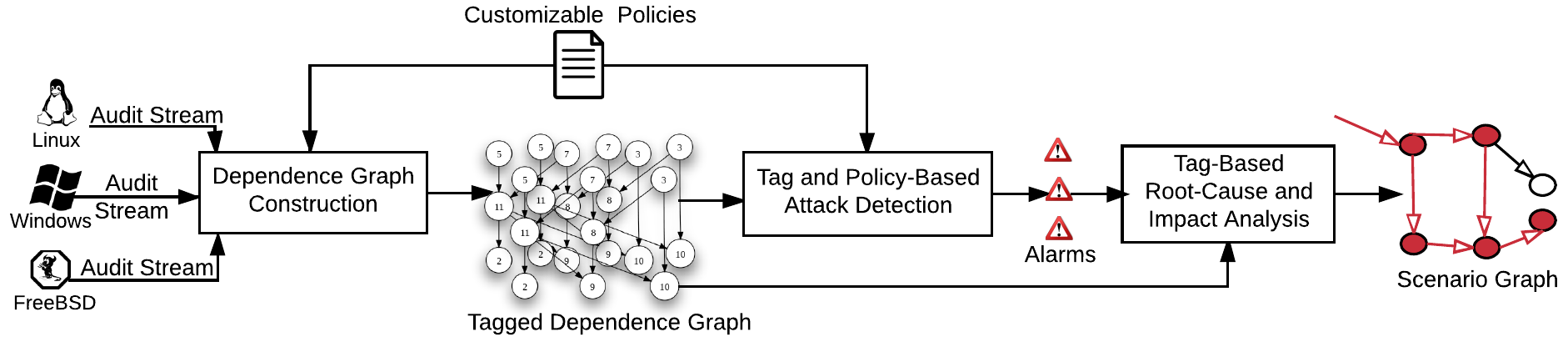}
    \caption{\projname System Overview \label{fig:architecture}
    }
\end{figure*} 

The problem of piecing together the causal chain of events leading to an attack
was first explored in Backtracker~\cite{king2003backtracking,king2005enriching}.
Subsequent research~\cite{lee2013high,ma2016protracer} improved on the precision
of the dependency chains constructed by Backtracker. However, these works
operate in a purely forensic setting and therefore do not deal with the
challenge of performing the analysis in real-time. In contrast, this paper
presents \projname,\footnote{\projname stands for (attack) Scenario LinkagE Using
  provenance Tracking of Host audit data.} a system that can alert analysts in
real-time about an ongoing campaign, and provide them with a compact, visual
summary of the activity in seconds or minutes after the attack. This would
enable a timely response before enormous damage is inflicted on the victim
enterprise.

Real-time attack detection and scenario reconstruction poses the
following additional challenges over a purely forensic analysis:
\begin{enumerate}
\item {\em Event storage and analysis:} How can we store the millions of records
  from event streams efficiently and have algorithms sift through this data in a
  matter of seconds?
\item {\em Prioritizing entities for analysis:} How can we assist the analyst,
  who is overwhelmed with the volume of data, prioritize and quickly ``zoom in'' on the most likely attack scenario?
\item {\em Scenario reconstruction:} How do we succinctly summarize the attack
  scenario, starting from the attacker's entry point and identifying the impact of
  the entire campaign on the system?
\item {\em Dealing with common usage scenarios:} How
  does one cope with normal, benign activities that may resemble activities
  commonly observed during attacks, e.g., software downloads? 
\item {\em Fast, interactive reasoning:} 
How can we provide the analyst with the ability to efficiently
  reason through the data, say, with an alternate hypothesis?
\end{enumerate}
Below, we provide a brief overview of \projname, and summarize our
contributions. \projname assumes that attacks initially come from outside the
enterprise. For example, an adversary could start the attack by hijacking a
web browser through externally supplied malicious input, by plugging in an
infected USB memory stick, or by supplying a zero-day exploit to a network
server running within the enterprise. We assume that the adversary {\em has not}
implanted persistent malware on the host {\em before} \projname started
monitoring the system. We also assume that the OS kernel and audit systems are
trustworthy.
\subsection{Approach Overview and Contributions}
Figure~\ref{fig:architecture} provides an overview of our approach. \projname is
OS-neutral, and currently supports Microsoft Windows, Linux and FreeBSD. Audit
data from these OSes is processed into a platform-neutral graph representation,
where vertices represent subjects (processes) and objects (files, sockets), and
edges denote audit events (e.g., operations such as read, write, execute, and
connect). This graph serves as the basis for attack detection as well as
causality analysis and scenario reconstruction.

The first contribution of this paper, which addresses the challenge of efficient
event storage and analysis, is the development of a compact main-memory
dependence graph representation (Section~\ref{sec:mainmem}). Graph algorithms on
main memory representation can be orders of magnitude faster than on-disk
representations, an important factor in achieving real-time analysis
capabilities. In our experiments, we were able to process 79 hours worth of
audit data from a FreeBSD system in 14 seconds, with a main memory usage of
84MB. This performance represents an analysis rate that is 20K times
faster than the rate at which the data was generated. 

The second major contribution of this paper is the development of a tag-based
approach for identifying subjects, objects and events that are most likely
involved in attacks. Tags enable us to prioritize and focus our analysis,
thereby addressing the second challenge mentioned above. Tags encode an
assessment of {\em trustworthiness} and {\em sensitivity} of data (i.e.,
objects) as well as processes (subjects). This assessment is based on data
provenance derived from audit logs. In this sense, tags derived from
audit data are similar to coarse-grain information flow labels. Our analysis can
naturally support finer-granularity tags as well, e.g., fine-grained taint tags
\cite{newsome05, usenix06}, if they are available. Tags are described in more
detail in Section~\ref{sec:detection}, together with their application to attack
detection.

A third contribution of this paper is the development of novel algorithms that
leverage tags for root-cause identification and impact analysis
(Section~\ref{sec:forensics}). Starting from alerts produced by the attack
detection component shown in Fig.~\ref{fig:architecture}, our backward analysis
algorithm follows the dependencies in the graph to identify the sources of the
attack.  Starting from the sources, we perform a full impact
analysis of the actions of the adversary using a forward search. We present
several criteria for pruning these searches in order to produce a compact graph.
We also present a number of transformations that further simplify this graph and produce a graph that visually captures the attack in a succinct and
semantically meaningful way, e.g., the graph in Fig.~\ref{fig:pandex}. Experiments show
that our tag-based approach is very effective: for instance, \projname can
analyze 38.5M events and produce an attack scenario graph with just 130 events,
representing five orders of magnitude reduction in event volume.

The fourth contribution of this paper, aimed at tackling the last two challenges
mentioned above, is a customizable policy framework (Section~\ref{sec:policy})
for tag initialization and propagation. Our framework comes with sensible
defaults, but they can be overridden to accommodate behaviors specific to an OS or application.  This enables tuning of our
detection and analysis techniques to avoid false positives in cases where
benign applications exhibit behaviors that resemble attacks. (See
Section~\ref{fabe} for details.) Policies also enable an analyst to test out
``alternate hypotheses'' of attacks, by reclassifying what is considered
trustworthy or sensitive and re-running the analysis. If an analyst suspects that some behavior is the
result of an attack, they can also use policies to capture these behaviors, and
rerun the analysis to discover its cause and impact. Since we can process and
analyze audit data tens of thousands of times faster than the rate at which it is
generated, efficient, parallel, real-time testing of alternate hypotheses is
possible.

The final contribution of this paper is an experimental evaluation
(Section~\ref{sec:eval}), based mainly on a red team evaluation organized by
DARPA as part of its Transparent Computing program. In this evaluation, attack
campaigns resembling modern APTs were carried out on Windows, FreeBSD and Linux
hosts over a two week period. In this evaluation, \projname was able to:
\begin{itemize}
\item process, in a matter of seconds, audit logs
containing tens of millions of events generated during the engagement;
\item successfully detect and reconstruct the details of these attacks, including
their entry points, activities in the system, and exfiltration points;
\item filter away extraneous events, achieving very high reductions rates in the 
data (up to 100K times), thus providing a clear semantic
representation of these attacks containing almost no noise from other activities
in the system; and 
\item achieve low false positive and false negative rates.
\end{itemize}
Our evaluation is not intended to show that we detected the most sophisticated
adversary; instead, our point is that, given several unknown possibilities, the
prioritized results from our system can be right on spot in real-time, without
any human assistance. Thus, it really fills a gap that exists today, where
forensic analysis seems to be primarily initiated manually.

\section{Main Memory Dependency Graph} \label{sec:mainmem}
To support fast detection and real-time analysis, we store dependencies in a
graph data structure. One possible option for storing this graph is a
graph database. However, the performance~\cite{mccoll2014performance} of popular
databases such as Neo4J~\cite{neo4j} or Titan~\cite{titan} is limited for many
graph algorithms unless main memory is large enough to hold most of data.
Moreover, the memory use of general graph databases is too high for our problem.
Even STINGER~\cite{ediger2012stinger} and NetworkX~\cite{networkx}, two graph
databases optimized for main-memory performance, use about 250 bytes
and 3KB, respectively, per graph edge \cite{mccoll2014performance}. The number
of audit events reported on enterprise networks can easily range in billions to
tens of billions per day, which will require main memory in the range of several
terabytes. In contrast, we present a much more
space-efficient dependence graph design that uses only about 10 bytes per edge. In one experiment, we
were able to store 38M events in just 329MB of main memory.

 The dependency graph is a per-host data structure. It can reference entities on
 other hosts but is optimized for the common case of intra-host reference. The
 graph represents two types of entities: {\em subjects}, which represent
 processes, and {\em objects}, which represent entities such as files, pipes,
 and network connections.
Subject attributes include process id (pid), command line, owner, and tags for
code and data. Objects attributes include name, type (file, pipe, socket, etc.),
owner, and tags. 

Events reported in the audit log are captured using labeled edges between
subjects and objects or between two subjects. For brevity, we use UNIX names such
as {\tt read}, {\tt connect}, and {\tt execve} for events.

We have developed a number of techniques to reduce storage requirements for
the dependence graph. Wherever possible, we use 32-bit identifiers instead
of 64-bit pointers. This allows a single host's dependence graph to contain
4 billion objects and subjects. The number of objects/subjects in our largest
data set was a few orders of magnitude smaller than this number.

While our design emphasizes compact data structures for objects and subjects,
compactness of events is far more important: events outnumber objects and
subjects by about two orders of magnitude in our largest data set. Moreover, the
ratio of events to objects+subjects increases with time. For this reason, we
have developed an ultra-compact representation for events that can use as little
as 6 bytes of storage for many events.

Events are stored inside subjects, thereby eliminating a need for
subject-to-event pointers, or the need for event identifiers. Their
representation uses variable-length encoding, so that in the typical case, they
can use just 4 bytes of storage, but when needed, they can use 8, 12, or 16
bytes. Most events operate on an object and have a timestamp. Since a
per-subject order of events is maintained, we dispense with microsecond
granularity for timestamps, instead opting for millisecond resolution. In
addition, we store only relative time since the last event on the same subject,
which allows us to do with 16-bits for the timestamp in the typical
case\footnote{Longer intervals are supported by recording a special ``timegap''
  event that can represent millions of years.}. Objects are referenced within
events using an index into a per-subject table of object identifiers. These
indices can be thought of like file descriptors --- they tend to have small
values, since most subjects use a relatively small number of objects. This
enables object references to be represented using 8 bits or less. We encode event names
for frequently occurring events (e.g., open, close, read and write) using 3 bits
or less. This leaves us with several bits for storing a summary
of event argument information, while still being within 32 bits.

We can navigate from subjects to objects using the event data stored within
subjects. However, forensic analysis also requires us to navigate from objects
to subjects. For this purpose, we need to maintain event information within
objects using object-event records. Object event records are maintained only
for a subset of events: specifically, events such as {\tt read} and {\tt write}
that result in a dataflow. Other events (e.g., {\tt open}) are not stored 
within objects. Object-event records are further shrunk by storing 
a reference to the corresponding subject-event record, instead of duplicating
information. 

As with subject-event records, we use a variable-length encoding for
object-event records that enables them to be stored in just 16 bits in the most
common case.  To see how this is possible, note that objects tend to be operated
on by a single subject at a time. Typically, this subject performs a sequence of
operations on the object, e.g., an open followed by a few reads or writes, and
then a close. By allowing object-event records to reuse the subject from their
predecessor, we can avoid the need for storing subject identifiers in most
records. Next, we allow object-event records to store a relative index for event
records within subjects. Two successive event records within a subject that
operate on the same object are likely to be relatively close to each other, say,
with tens or hundreds of events in-between. This means that the relative index
stored with object-event record can be 12 bits or less in most cases, thus
allowing these records to be 16 bits or less in the typical case.

This design thus allows us to store bidirectional timestamped edges
in as little as 6 bytes (4 bytes for a subject-event record and 2 bytes for an object-event record).  In experiments with larger data sets, the total memory use of our system is within 10 bytes per event on average.

Our variable length encoding allows us to represent full
information about important (but rare) events, such as rename, chmod, execve,
and so on. So, compactness is achieved without losing any important
information. Although such encoding slows down access, 
access times are still typically less than 100ns, which is many orders of
magnitude faster than disk latencies that dominate random access on 
disk-resident data structures.

\section{Tags and Attack Detection}
\label{sec:detection}
We use tags to summarize our assessment of the trustworthiness and sensitivity
of objects and subjects. This assessment can be based on three main factors:
\begin{itemize}
\item {\em Provenance:} the tags on the immediate predecessors of an
  object or subject in the dependence graph,
\item {\em Prior system knowledge:} our knowledge about the behavior
  of important applications, such as remote access servers and software
   installers, and important files such as \mytt{/etc/passwd} and
   \mytt{/dev/audio}, and
\item {\em Behavior:} observed behavior of subjects, and how they
  compare to their expected behavior.
\end{itemize}
We have developed a policy framework, described in Section~\ref{sec:policy}, for
initializing and propagating tags based on these factors. In the absence of
specific policies, a default policy is used that propagates tags from inputs to
outputs. The default policy assigns to an output the lowest among the
trustworthiness tags of the inputs, and the highest among the confidentiality tags.
This policy is conservative: it can err on the side of over-tainting,
but will not cause attacks to go undetected, or cause a forward (or backward)
analysis to miss objects, subjects or events.

Tags play a central role in \projname. They provide important context for attack
detection.  Each audited event is interpreted in the context of
these tags to determine its likelihood of contributing to an attack. In
addition, tags are instrumental for the speed of our forward and backward
analysis. Finally, tags play a central role in scenario reconstruction by
eliminating vast amounts of audit data that satisfy the technical
definition of dependence but do not meaningfully contribute to our understanding
of an attack. 
\subsection{Tag Design}
\label{subsec:tags}
We define the following {\em trustworthiness tags} ({\em t-tags}):
\begin{itemize}
\item {\em Benign authentic} tag is assigned to data/code received from sources
  trusted to be benign, and whose authenticity can be verified.
\item {\em Benign} tag reflects a reduced level of trust than benign authentic:
  while the data/code is still believed to be benign, adequate authentication hasn't
  been performed to verify the source.
\item {\em Unknown} tag is given to data/code from sources
  about which we have no information on trustworthiness. Such data {\em
    can sometimes be} malicious.
\end{itemize}
Policies define what sources are benign and what forms of authentication are
sufficient. In the simplest case, these policies take the form of whitelists,
but we support more complex policies as well. If no policy is applicable to a
source, then its t-tag is set to {\em unknown.}

We define the following {\em confidentiality tags} ({\em c-tags}), 
to reason about information stealing attacks:
\begin{itemize}
\item {\em Secret:} Highly sensitive information, such
  as login credentials and private keys.
\item {\em Sensitive:} Data whose disclosure can have a significant
  security impact, e.g., reveal vulnerabilities in the system,
  but does not provide a direct way for an attacker to gain
  access to the system.
\item {\em Private:} Data whose disclosure is a privacy concern, but
  does not necessarily pose a security threat.
\item {\em Public:} Data that can be widely available, e.g., on
  public web sites.
\end{itemize}

An important aspect of our design is the separation between t-tags for code and
data. Specifically, a subject (i.e., a process) is given two t-tags: one that
captures its {\em code trustworthiness} (code t-tag) and another for its {\em
  data trustworthiness} (data t-tag). This separation significantly improves
attack detection. More importantly, it can significantly speed up forensic
analysis by focusing it on fewer suspicious events, while substantially reducing
the size of the reconstructed scenario. Note that confidentiality tags are
associated only with data (and not code).

Pre-existing objects and subjects are assigned initial tags using {\em tag
  initialization policies.} Objects representing external entities, such as a
remote network connection, also need to be assigned initial tags. The rest of
the objects and subjects are created during system execution, and their tags are
determined using {\em tag propagation policies.} Finally, attacks are detected
using behavior-based policies called {\em detection policies.}

As mentioned before, if no specific policy is provided, then sources are
tagged with {\em unknown} trustworthiness. Similarly, in the absence of specific propagation
 policies, the default conservative propagation policy is used.

\subsection{Tag-based Attack Detection}
\comment{
Numerous exploit detection and prevention techniques have been developed over
the past two decades, including techniques such as memory layout and code
randomization \cite{...}, control-flow integrity \cite{...}, information flow
techniques \cite{...}, sandboxing \cite{...}, privilege separation (aka broker
architecture by vendors) \cite{...}, and so on. Even though these techniques are
secured by custom instrumentation embedded into protected systems, attackers
continue to find ways to bypass or evade these techniques. Since \projname
operates without any active instrumentation, we need a different strategy and
approach for attack detection, or else it will be far to easy to evade. It is
this factor that motivated the development of our tag-based detection technique.
}
\comment{
  We therefore use provenance to drive attack detection. We define broad,
application-independent policies on the use of data based on its provenance,
with policy violations raising alerts that initiate a forensic analysis.
\todo{fix this -- soften the text}
Almost every reliable detection method available today is focused on the
specifics of exploit mechanisms used, such as memory corruption, control-flow
hijack, SQL injection, and so on. This factor limits these methods to narrow
classes of attacks. Resourceful adversaries are often able to craft
attacks that exploit mechanisms for which no protection measures
have been deployed on the victim system, e.g., sandbox escape attacks,
or native code execution via spearphishing.
}
An important constraint in \projname is that we are limited to information
available in audit data. This suggests the use of provenance
reflected in audit data as a possible basis for detection. Since tags are a
function of provenance, we use them for attack detection. Note that
in our threat model, audit data is trustworthy, so tags provide a sound basis
for detection.

A second constraint in \projname is that detection methods should not require
detailed application-specific knowledge. In contrast, most existing intrusion
detection and sandboxing techniques interpret each security-sensitive operation
in the context of a specific application to determine whether it could be malicious.
This requires expert knowledge about the application, or in-the-field training
in a dynamic environment, where applications may be frequently updated.

Instead of focusing on application behaviors that tend to be variable, we focus
our detection techniques on the high-level objectives of most attackers, such as
backdoor insertion and data exfiltration. Specifically, we combine reasoning
about an attacker's {\em motive} and {\em means.} If an event in the audit data can help the attacker achieve his/her key high-level
objectives, that would provide the motivation and justification for using that
event in an attack. But this is not enough: the attacker also needs the means to
cause this event, or more broadly, influence it. Note that our tags are
designed to capture means: if a piece of data or code bears the {\em unknown}
\mbox{t-tag}, then it was derived from (and hence influenced by)
untrusted sources.

As for the high-level objectives of an attacker, several reports and white
papers have identified that the following steps are typical in most advanced
attack campaigns~\cite{apt-reports,killchain,mandiant-report}:
\begin{enumerate}
\item Deploy and run attacker's code on victim system.
\item Replace or modify important files, e.g., {\tt /etc/passwd} or
  ssh keys.%
\item Exfiltrate sensitive data.
\end{enumerate}
Attacks with a transient effect may be able to avoid the first two steps, but
most sophisticated attacks, such as those used in APT campaigns, require the
establishment of a more permanent footprint on the victim system. In those
cases, there does not seem to be a way to avoid one or both of the first two
steps. Even in those cases where the attacker's goal could be achieved
without establishing a permanent base, the third step usually represents an
essential attacker goal.

Based on the above reasoning, we define the following policies for attack
detection that incorporate the attacker's objectives and means:
\begin{itemize}
\item {\em Untrusted code execution:} This policy triggers an alarm when a
  subject with a higher code t-tag executes (or loads) an object with a lower
  t-tag\footnote{Customized policies can be defined for interpreters such as
    {\tt bash} so that reads are treated the same as loads.}.
\item {\em Modification by subjects with lower code t-tag:} This policy raises
  an alarm when a subject with a lower code t-tag modifies an object with a
  higher t-tag. Modification may pertain to the file content or other
  attributes such as name, permissions, etc.
\item {\em Confidential data leak:} An alarm is raised when untrusted subjects
  exfiltrate sensitive data. Specifically, this policy is triggered on
  network writes by subjects with a {\em sensitive} c-tag and a code t-tag of
  {\em unknown.}
\item {\em Preparation of untrusted data for execution:} This policy is
  triggered by an operation by a subject with a code t-tag of {\em unknown,}
  provided this operation makes an object executable. Such operations include
  {\tt chmod} and {\tt mprotect}\footnote{Binary code injection
    attacks on today's OSes ultimately involve a call to change the permission
    of a writable memory page so that it becomes executable. To the extent that
    such memory permission change operations are included in the audit data,
    this policy can spot them.}$^,$\footnote{Our implementation can
    identify {\tt mprotect} operations that occur in conjunction with library
    loading operations. This policy is not triggered on those {\tt
      mprotect}'s.}.
  \comment{
    Leave the following out to conserve space, plus it does not play well with
    the Steiner tree formulation for backward analysis.
    
\item {\em Subject downgrade:} This policy is triggered when a subject's
  data t-tag is lowered because of reading a lower integrity object.
  Most subjects do not expect to handle untrusted content, and may be
  easily exploited by them. This is especially true when you consider that
  some of these files may be configuration files rather than data files.
  Customized policies are used to exempt some applications that do expect
  to handle untrusted inputs, e.g., network servers or
  web browsers.
  }
\end{itemize}
It is important to note that ``means'' is not diluted just because data or code
passes through multiple intermediaries. For instance, the untrusted code policy
does not require a direct load of data from an unknown web site; instead, the
data could be downloaded, extracted, uncompressed, and possibly compiled, and
then loaded. Regardless of the number of intermediate steps, this policy will be
triggered when the resulting file is loaded or executed. This is one of the
most important reasons for the effectiveness of our attack detection. 

Today's vulnerability exploits typically do not involve untrusted code in their
first step, and hence won't be detected by the untrusted code execution policy.
However, the eventual goal of an attacker is to execute his/her code, either by
downloading and executing a file, or by adding execute permissions to a memory
page containing untrusted data. In either case, one of the above policies can
detect the attack. A subsequent backward analysis can help identify the first
step of the exploit.

Additional detector inputs can be easily integrated
into \projname. For instance, if an external detector flags a subject as a
suspect, this can be incorporated by setting the code t-tag of the subject to 
{\em unknown.} As a result, the remaining detection policies mentioned above can
all benefit from the information provided by the external detector. Moreover,
setting of {\em unknown} t-tag at suspect nodes preserves the dependency
structure between the graph vertices that cause alarms, a fact that we
exploit in our forensic analysis. 

The fact that many of our policies are triggered by untrusted code execution
should not be interpreted to mean that they work in a static environment,
where no new code is permitted in the system. Indeed, we expect software
updates and upgrades to be happening constantly, but in an enterprise
setting, we don't expect end users to be downloading unknown code from
random sites. Accordingly, we subsequently describe how to support
standardized software updating mechanisms such as those used on contemporary
OSes.

\section{Policy Framework}
\label{sec:policy}
We have developed a flexible policy framework for tag assignment, propagation,
and attack detection. We express policies using a simple rule-based notation,
e.g., %
\[
\small exec(s, o)\!:\; o.ttag < benign \rightarrow alert(\mbox{\tt
    "UntrustedExec"})%
\]
This rule is triggered when the subject $s$ executes a (file) object $o$ with a
t-tag less than $benign$. Its effect is to raise an alert named {\tt
  UntrustedExec}. As illustrated by this example, rules are generally associated
with events, and include conditions on the attributes of objects and/or subjects
involved in the event. Attributes of interest include:
\begin{itemize}
\item {\em name:} regular expressions can be used to match object names
  and subject command lines. We use Perl syntax for regular expressions.
\item {\em tags:} conditions can be placed on t-tags and c-tags of
  objects and/or subjects. For subjects, code and data t-tags can be
  independently accessed. 
\item {\em ownership and permission:} conditions can be placed on
  the ownership of objects and subjects, or permissions associated with 
  the object or the event.
\end{itemize}
The effect of a policy depends on its type. The effect of a detection
policy is to raise an alarm. For tag initialization and propagation
policies, the effect is to modify tag(s) associated with the object or
subject involved in the event. While we use a rule-based notation to specify
policies in this paper, in our implementation, each rule is encoded as a (C++)
function.

To provide a finer degree of control over the order in which different types of
policies are checked, we associate policies with {\em trigger points}
instead of events. In addition, trigger points provide a level of indirection
that enables sharing of policies across distinct events that have a similar
purpose. Table~\ref{fig:event-type} shows the trigger points currently
defined in our policy framework. The first column identifies events, the
second column specifies the direction of information flow, and the last
two columns define the trigger points associated with these events.  

\newcommand{\ra}{\rightarrow}
\begin{table}
  \begin{center}
    \begin{adjustbox}{width=0.9\columnwidth}
       \begin{tabular}{||c|c|c|c||}
       \hline
       Event   & Direction & Alarm & Tag \\
               &           & trigger& trigger \\
       \hline
       {\tt define} & & & $init$ \\
       \hline
       {\tt read}   & O$\ra$S & $read$         & $propRd$\\
       \hline
       {\tt load}, {\tt execve}   & O$\ra$S & $exec$ & $propEx$\\
       \hline
       {\tt write}  & S$\ra$O & $write$         & $propWr$\\
       \hline
       {\tt rm}, {\tt rename}     & S$\ra$O & $write$       &      \\
       \hline
       {\tt chmod}, {\tt chown}  & S$\ra$O & $write$, $modify$  &      \\
       \hline
       {\tt setuid} & S$\ra$S &            & $propSu$\\
       \hline
       \end{tabular}
       \end{adjustbox}
    \end{center}
  \caption{Edges with policy trigger points. In the direction column, S
    indicates subject, and O indicates object. The next two columns indicate
    trigger points for detection policies and tag setting policies.}
  \label{fig:event-type}
\end{table}

Note that we use a special event called {\tt define} to denote audit records
that define a new object. This pseudo-event is assumed to have occurred when a
new object is encountered for the first time, e.g., establishment of a new
network connection, the first mention of a pre-existing file, creation of a new
file, etc. The remaining events in the table are self-explanatory.

When an event occurs, all detection policies associated with its alarm trigger are
executed. Unless specifically configured, detection policies are checked only
when the tag of the target subject or object is about to change. (``Target''
here refers to the destination of data flow in an operation.) Following this,
policies associated with the event's tag triggers are tried in the order in
which they are specified. As soon as a matching rule is found, the tags
specified by this rule are assigned to the target of the event, and the
remaining tag policies are not evaluated.

Our current detection policies are informally described in the previous
section. We therefore focus in this section on our current 
tag initialization and propagation policies.
\subsection{Tag Initialization Policies}
These policies are invoked at the $init$ trigger, and are used to initialize
tags for new objects, or preexisting objects when they are first mentioned in
the audit data. Recall that when a subject creates a new object, the object
inherits the subject's tags by default; however, this can be overridden
using tag initialization policies. 

Our current tag initialization policy is as follows. Note the use of regular
expressions to conveniently define initial tags for groups of objects.\\[0.8ex]
\indent{\small
$init(o)$: $match(o.name, \verb+"^IP:(10\.0|127)"+) \rightarrow \\
\hsp o.ttag={\tt BENIGN\_AUTH}, o.ctag={\tt PRIVATE}$

$init(o)$: $match(o.name, \verb+"^IP:"+) \rightarrow \\
\hsp o.ttag={\tt UNKNOWN}, o.ctag={\tt PRIVATE}$

$init(o)$: $o.type == {\tt FILE} \rightarrow \\
\hsp o.ttag={\tt BENIGN\_AUTH}, o.ctag={\tt PUBLIC}$}\\[0.8ex]
The first rule specifies tags for intranet connections, identified by address
prefixes 10.0 and 127 for the remote host. It is useful in a context where
\projname isn't deployed on the remote host\footnote{If \projname is deployed on
  the remote host, there will be no {\tt define} event associated with the
  establishment of a network connection, and hence this policy won't be
  triggered. Instead, we will already have computed a tag for the remote network
  endpoint, which will now propagate to any local subject that reads from the
  connection.}. The second rule states that all other hosts are untrusted.
All preexisting files are assigned the same tags by the third rule.
Our implementation uses two additional policies that specify c-tags. 
\subsection{Tag Propagation Policies}
These policies can be used to override default tag propagation semantics.
Different tag propagation policies can be defined for different groups of
related event types, as indicated in the ``Tag trigger'' column in Table
\ref{fig:event-type}.

Tag propagation policies can be used to prevent ``over-tainting'' that can
result from files such as {\tt .bash\_history} that are repeatedly read and
written by an application each time it is invoked. The following policy
skips taint propagation for this specific file: \\[0.8ex]
{\small $propRd(s, o)$: $match(o.name, \verb+"\.bash_history$"+) \rightarrow
  skip\footnote{Here, ``skip'' means do nothing, i.e., leave tags unchanged.}$}\\[0.8ex]
\noindent Here is a policy that treats files read by {\tt bash}, which is
an interpreter, as a load, and hence updates the code t-tag.\\[0.8ex]
{\small
  $propRd(s, o)$: $match(s.cmdline, \verb+"^/bin/bash$"+) \rightarrow\\
  \hsp s.code\_ttag=s.data\_ttag=o.ttag, s.ctag=o.ctag$\\[0.8ex]
}
Although trusted servers such as {\tt sshd} interact with untrusted
sites, they can be expected to protect themselves, and let only authorized users
access the system. Such servers should not have their data trustworthiness
downgraded.  A similar comment applies to programs such as software updaters and
installers that download code from untrusted sites, but verify the signature of
a trusted software provider before the install.\\[0.8ex]
\noindent {\small $propRd(o, s)$: $match(s.cmdline, \verb+"^/sbin/sshd$"+) \rightarrow skip$}\\[0.8ex]
Moreover, when the login phase is complete, typically
identified by execution of a {\tt setuid} operation, the process should
be assigned appropriate tags.\\[0.8ex]
{\small
  $propSu(s)$: $match(s.cmdline, \verb+"^/usr/sbin/sshd$"+) \rightarrow\\
  \hsp s.code\_ttag=s.data\_ttag={\tt BENIGN}, s.ctag={\tt PRIVATE}$%
}

\section{Tag-Based Bi-Directional Analysis}
\label{sec:forensics}
\subsection{Backward Analysis}
The goal of backward analysis is to identify the entry points of an attack
campaign. Entry points are the nodes in the graph with an in-degree of zero and
are marked untrusted. Typically they represent network connections, but they can
also be of other types, e.g., a file on a USB stick that was plugged into the
victim host.

The starting points for the backward analysis are the alarms generated by the
detection policies. In particular, each alarm is related to one or more entities, which
are marked as suspect nodes in the graph. Backward search involves a backward
traversal of the graph to identify paths that connect the suspect nodes to
entry nodes. We note that the direction of the dependency edges is reversed in
such a traversal and in the following discussions. Backward search poses
several significant challenges:
\begin{itemize}
\item {\em Performance:} The dependence graph can easily contain hundreds
  of millions of edges. Alarms can easily number in
  thousands. Running backward searches on such a large graph is computationally
  expensive.
\item {\em Multiple paths:} Typically numerous entry points are backward reachable
  from a suspect node. However, in APT-style attacks, there is often just one
  real entry point. Thus, a naive backward search can lead to a large number of 
  false positives. 
\end{itemize}
The key insight behind our approach is that tags can be used to address both
challenges. In fact, tag computation and propagation is already an implicit
path computation, which can be reused. Furthermore, a tag value of {\em unknown}
on a node provides an important clue about the likelihood of that node being a
potential part of an attack. In particular, if an {\em unknown} tag exists for
some node $A$, that means that there exists at least a path from an untrusted
entry node to node $A$, therefore node $A$ is more likely to be part of an
attack than other neighbors with {\em benign} tags. Utilizing tags for the
backward search greatly reduces the search space by eliminating many irrelevant
nodes and sets \projname apart from other scenario reconstruction approaches
such as~\cite{king2003backtracking,lee2013high}.

Based on this insight, we formulate backward analyis as an instance of shortest
path problem, where tags are used to define edge costs. In effect, tags are able
to ``guide'' the search along relevant paths, and away from unlikely paths. This
factor enables the search to be completed without necessarily traversing the
entire graph, thus addressing the performance challenge. In addition, our
shortest path formulation addresses the multiple paths chalenge by by preferring
the entry point closest (as measured by path cost) to a suspect node.

For shortest path, we use Dijkstra's algorithm, as it discovers paths
in increasing order of cost. In particular, each step of this algorithm 
adds a node to the shortest path tree, which consists of the shortest
paths computed so far. This enables the search to stop as soon as an
entry point node is added to this tree. 

\paragraph{Cost function design.}
 Our design assigns low costs to edges representing dependencies on nodes with
 {\em unknown} tags, and higher costs to other edges. Specifically, the costs
 are as follows:

\begin{itemize}
\item Edges that introduce a dependency from a node with {\em unknown} code or
  data t-tag to a node with {\em benign} code or data t-tag are assigned a cost
  of 0.
\item Edges introducing a dependency from a node with {\em benign} code 
  and data t-tags are assigned a high cost.
\item Edges introducing dependencies between nodes already having
  an {\em unknown} tag are assigned a~cost~of~1.
\end{itemize}
The intuition behind this design is as follows. A benign subject or object
immediately related to an {\tt unknown} subject/object represents the boundary
between the malicious and benign portions of the graph. Therefore, they must be
included in the search, thus the cost of these edges is 0. Information flows
among benign entities are not part of the attack, therefore we set their cost to
very high so that they are excluded from the search. Information flows among
untrusted nodes are likely part of an attack, so we set their cost to a low
value. They will be included in the search result unless alternative
paths consisting of fewer edges are available.

\begin{table*}
\begin{center}
\begin{footnotesize}
\begin{tabular}{||c|c|c|c|c|c|c|c|c|c|c|c||}
\hline
Dataset & \begin{tabular}[c]{@{}c@{}}Duration \\(hh-mm-ss) \end{tabular} & Open &\begin{tabular}[c]{@{}c@{}} Connect + \\Accept \end{tabular} & Read    & Write  &\begin{tabular}[c]{@{}c@{}} Clone + \\Exec \end{tabular}&\begin{tabular}[c]{@{}c@{}} Close + \\Exit \end{tabular}&\begin{tabular}[c]{@{}c@{}} Mmap /\\ Loadlib  \end{tabular} & Others &\begin{tabular}[c]{@{}c@{}} Total \# of\\Events \end{tabular}& \begin{tabular}[c]{@{}c@{}}Scenario\\Graph\end{tabular}\\ \hline \hline
W-1       & 06:22:42 &  N/A	& 22.14\%	&44.70\%	&5.12\% 	&3.73\%	& 3.88\%	&17.40\%	&3.02\%   & 100K       &  Fig. \ref{5d-stretch} \\ \hline
W-2       & 19:43:46 &  N/A	& 17.40\%	&47.63\%	&8.03\% 	&3.28\%	& 3.26\%	&15.22\%	&5.17\%   & 401K       &  Fig. \ref{fdbov1} \\ \hline \hline
L-1       & 07:59:26 & 37\%	& 0.11\%	&18.01\%	&1.15\% 	&0.92\%	&38.76\%	& 3.97\%	&0.07\%   & 2.68M      & Fig. \ref{sristr_2} \\ \hline
L-2       & 79:06:39 & 39.58\%	& 0.08\%	&12.19\%	&2\%	        &0.83\%	&41.28\%	& 3.79\%	&0.25\%   & 38.5M      & -\\ \hline
L-3       & 79:05:13 & 38.88\%	& 0.04\%	&11.81\%	&2.35\% 	&0.95\%	&40.98\%	& 4.14\%	&0.84\%   & 19.3M      & Fig. \ref{sri-pandex}\\ \hline \hline
F-1       & 08:17:30 & 9.46\%	& 0.40\%	&24.65\%	&40.86\%	&2.10\%	&12.55\%	& 9.08\%	&0.89\%   & 701K       & Fig. \ref{cadets-stretch} \\ \hline
F-2       & 78:56:48 & 11.78\%	& 0.42\%	&16.60\%	&44.52\%	&2.10\%	&15.04\%	& 8.54\%	&1.01\%   & 5.86M      & Fig. \ref{cadets-bovia}    \\ \hline
F-3       & 79:04:54 & 11.31\%	& 0.40\%	&19.46\%	&45.71\%	&1.64\%	&14.30\%	& 6.16\%	&1.03\%   & 5.68M      & Fig. \ref{fig:pandex}     \\ \hline \hline
Benign    &329:11:40 & 11.68\%	& 0.71\%	&26.22\%	&30.03\%	&0.63\%	&15.42\%	&14.32\%	&0.99\%   & 32.83M     & N/A\\ \hline
\end{tabular}
\end{footnotesize}
\end{center}
\caption{Dataset for each campaign with duration, distribution of different system calls and total number of events. } \label{tab:dist}
\end{table*}

\subsection{Forward Analysis}
The purpose of forward analysis is to assess the impact of a campaign, by starting
from an entry point and discovering all the possible effects dependent on the
entry point. Similar to backward analysis, the main challenge is
the size of the graph. A naive approach would identify and flag all  subjects
and objects reachable from the entry point(s) identified by backward
analysis. Unfortunately, such an approach will result in an impact graph that is
too large to be useful to an analyst. For instance, in our experiments, a naive
analysis produced impact graphs with millions of edges, whereas our refined
algorithm reduces this number by 100x to 500x.

A natural approach for reducing the size is to use a
distance threshold $d_{th}$ to exclude nodes that are ``too far'' from the
suspect nodes. Threshold $d_{th}$ can be interactively tuned by an analyst. We
use the same cost metric that was used for backward analysis, but modified to
consider confidentiality\footnote{Recall that some alarms are related to
  exfiltration of confidential data, so we need to decide which edges
  representing the flow of confidential information should be included in the
  scenario.}. In particular, edges between nodes with high confidentiality tags
(e.g., {\em secret}) and nodes with low code integrity tags (e.g., {\em unknown}
process) or low data integrity tags (e.g., {\em unknown} socket) are assigned a
cost of 0, while edges to nodes with {\em benign} tags are assigned a high cost.

\comment{
If the graph identified by the above analysis is still too large, 
another refinement is performed. In particular, rather than
following all the paths from the entry node, we follow paths along nodes that
are at most $d_{th}$ edges far from the nodes in the path computed by backward
analysis, where $d_{th}$ is a threshold that can be set by the analyst. This
approach can significantly trim the sizes of graphs in cases where ``phantom''
dependencies result in a large number of nodes being marked with lower t-tags.
}

\subsection{Reconstruction and Presentation}\label{simplifications}
\label{sec:simplifications}
We apply the following simplifications to the output of forward analysis,
in order to provide a more succinct view of the attack:
\begin{itemize}
\item {\em Pruning uninteresting nodes.} The result of forward analysis may
include many dependencies that are not relevant for the attack, e.g., subjects
writing to cache and log files, or writing to a temporary file and then
removing it. These nodes may appear
in the results of the forward analysis but no suspect nodes depend on them,
so they can be pruned.
\item {\em Merging entities with the same name.}
This simplification merges subjects that have the same name, disregarding their
process ids and command-line arguments.
\item {\em Repeated event filtering.} 
This simplification merges into one those events that happen
multiple times (e.g., multiple writes, multiple reads) between the same
entities. If there are interleaving events, then we show two events representing
the first and the last occurrence of an event between the two entities.
\end{itemize}

\section{Experimental Evaluation}
\label{sec:eval}
\subsection{Implementation}
Most components of \projname, including the graph model, policy
engine, attack detection and some parts of the forensic analysis are implemented
in C++, and consist of about 9.5KLoC. The remaining components, including
that for reconstruction and presentation, are implemented in Python,
and consist of 1.6KLoC. 
\subsection{Data Sets}
Table \ref{tab:dist} summarizes the dataset used in our evaluation. The first
eight rows of the table correspond to attack campaigns carried out by a red team
as part of the DARPA Transparent Computing (TC) program. This set spans a period
of 358 hours, and contains about 73 million events. The last row corresponds to
benign data collected over a period of 3 to 5 days across four Linux servers in
our research laboratory.

Attack data sets were collected on Windows (W-1 and W-2), Linux (L-1 through
L-3) and FreeBSD (F-1 through F-3) by three research teams that are also part of
the DARPA TC program. The goal of these research teams is to provide
fine-grained provenance information that goes far beyond what is found in
typical audit data. However, at the time of the evaluation, these advanced
features had not been implemented in the Windows and FreeBSD data sets. Linux
data set did incorporate finer-granularity provenance (using the unit
abstraction developed in \cite{lee2013high}), but the implementation was not
mature enough to provide consistent results in our tests. For this reason, we
omitted any fine-grained provenance included in their dataset, falling back to
the data they collected from the built-in auditing system of Linux. The FreeBSD
team built their capabilities over DTrace. Their data also corresponded to
roughly the same level as Linux audit logs. The Windows team's data was roughly
at the level of Windows event logs. All of the teams converted their data into a
common representation to facilitate analysis.

The ``duration" column in Table \ref{tab:dist} refers to the length of time for
which audit data was emitted from a host. Note that this period covers both
benign activities and attack related activities on a host. The next several
columns provide a break down of audit log events into different types of
operations. File open and close operations were not included in W-1 and W-2 data
sets. Note that ``read'' and ``write'' columns include not only file
reads/writes, but also network reads and writes on Linux. However, on Windows,
only file reads and writes were reported. Operations to load libraries were
reported on Windows, but memory mapping operations weren't. On Linux and FreeBSD,
there are no load operations, but most of the mmap calls are related to
loading. So, the mmap count is a loose approximation of the number of loads on
these two OSes. The ``Others'' column includes all the remaining audit
operations, including {\tt rename}, {\tt link}, {\tt rm}, {\tt unlink}, {\tt
  chmod}, {\tt setuid}, and so on. The last column in the table identifies the
scenario graph constructed by \projname for each campaign. Due to space
limitations, we have omitted scenario graphs for campaign L-2.

 \newcommand\ts{\rule{0pt}{2.1ex}}         %
 \newcommand\bs{\rule[-2.9ex]{0pt}{0pt}}   %

\begin{figure*}
  \begin{center}
    \includegraphics[width=5.7in,height=2.9in]{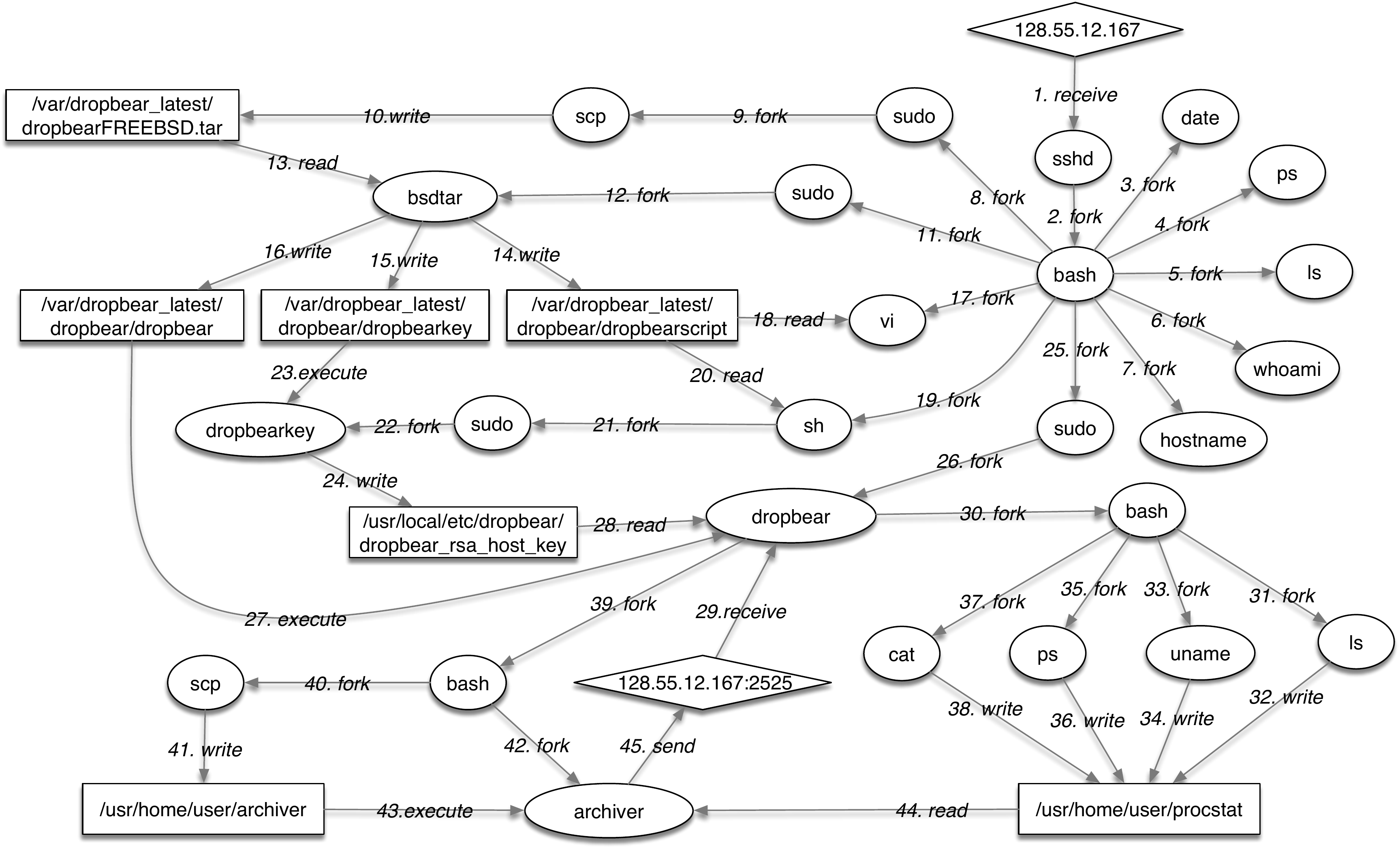}
  \end{center}
    \caption{Scenario graph reconstructed from campaign F-3.}
    \label{fig:pandex}
\end{figure*} 

\subsection{Engagement Setup}

 The attack scenarios in our evaluation are setup as follows. Five of the
 campaigns (i.e., W-2, L-2, L3, F-2, and F3) ran in parallel for 4 days, while
 the remaining three (W-1, L-1, and F-1) were run in parallel for 2 days. During
 each campaign, the red team carried out a series of attacks on the target
 hosts. The campaigns are aimed at achieving varying adversarial objectives,
 which include dropping and execution of an executable, gathering intelligence
 about a target host, backdoor injection, privilege escalation, and data
 exfiltration.

 Being an adversarial engagement, we had no prior knowledge of the attacks
 planned by the red team. We were only told the broad range of attacker
 objectives described in the previous paragraph. It is worth noting that, while
 the red team was carrying out attacks on the target hosts, benign background
 activities were also being carried out on the hosts. These include activities
 such as browsing and downloading files, reading and writing emails, document
 processing, and so on. On average, {\em more than 99.9\% of the events
   corresponded to benign activity.} Hence, \projname had to automatically
 detect and reconstruct the attacks from a set of events including both benign
 and malicious activities.
 
We present our results in comparison with the ground truth data released by the
red team. Before the release of ground truth data, we had to provide a report of
our findings to the red team. The findings we report in this paper match the
findings we submitted to the red team. A summary of our detection and
reconstruction results is provided in a tabular form in
Table~\ref{detection-summary}. Below, we first present reconstructed scenarios
for selected datasets before proceeding to a discussion of these summary
results.

\subsection{Selected Reconstruction Results}\label{sec:selres}
Of the 8 attack scenarios successfully reconstructed by \projname, we discuss
campaigns W-2 (Windows) and F-3 (FreeBSD) in this section, while deferring the
rest to Section \ref{subsec:additional}. To make it easier to follow the
scenario graph, we provide a narrative that explains how the attack unfolded.
This narrative requires manual interpretation of the graph, but the graph
generation itself is automated. In these graphs, edge labels include the event
name and a sequence number that indicates the global order in which that event
was performed. Ovals, diamonds and rectangles represent processes,
sockets and files, respectively. 
\paragraph{Campaign W-2.}
\begin{figure}[b]
\begin{center}
\vspace*{-1.3em}
\includegraphics[scale=0.35]{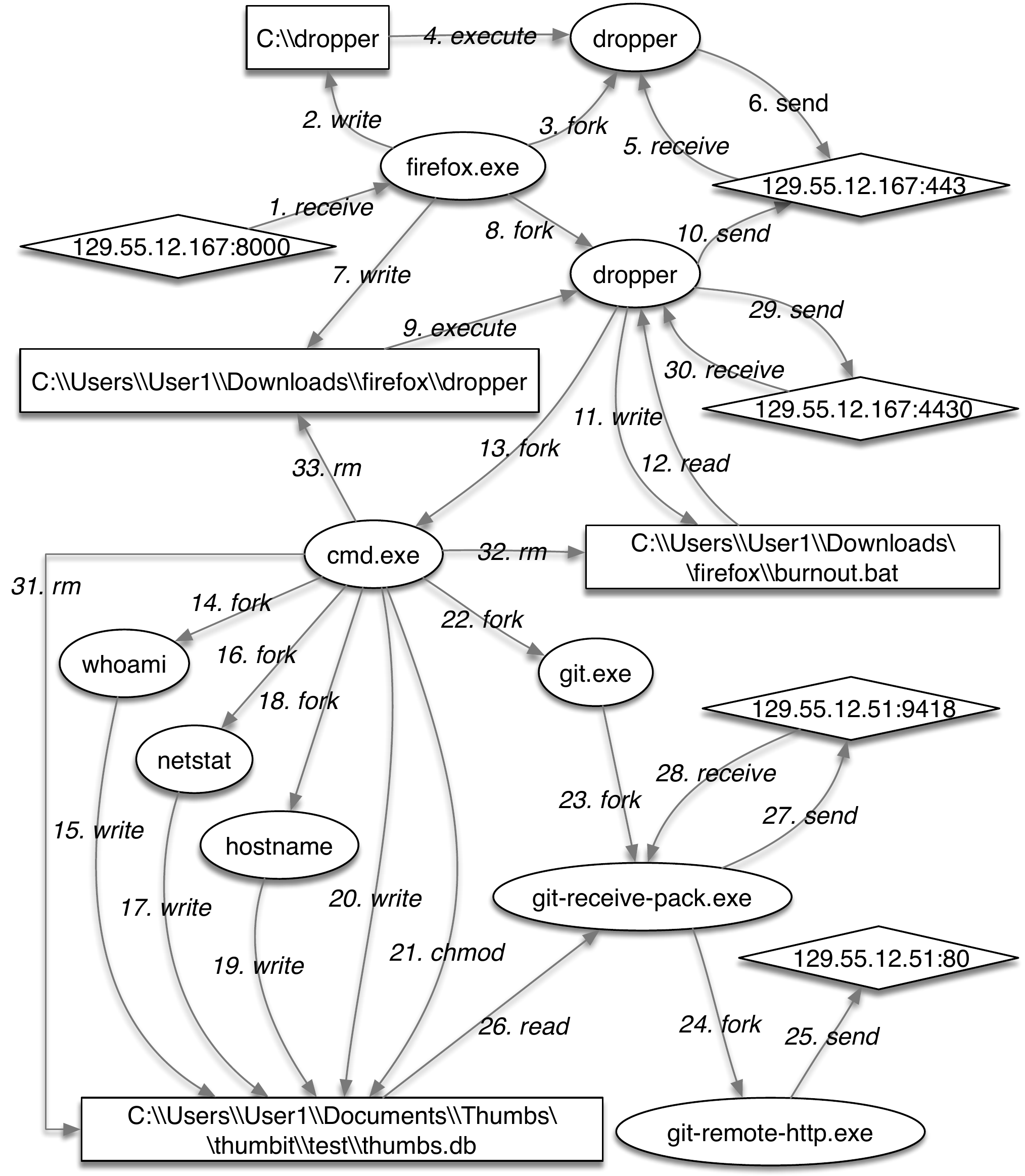}
\end{center}
\caption{ Scenario graph reconstructed from campaign W-2.}
\label{fdbov1}
\end{figure}
Figure~\ref{fdbov1} shows the graph reconstructed by \projname from Windows
audit data. Although the actual attack campaign lasted half an hour, the
host was running benign background activities for 20 hours. These
background activities corresponded to more than 99.8\% of the events in
the corresponding audit log.

\noindent {\em Entry}: The initial entry point for the attack is {\tt Firefox},
which is compromised on visiting the web server {\tt 129.55.12.167}.

\noindent {\em Backdoor insertion}: Once Firefox is compromised, a malicious
program called {\tt dropper} is downloaded and executed. Dropper seems to 
provide a remote interactive shell, connecting to ports 443 and then 4430 on
the attack host, and executing received commands using {\tt cmd.exe}.

\noindent {\em Intelligence gathering}: Dropper then invokes {\tt cmd.exe}
multiple times, using it to perform various data gathering tasks. The programs
{\tt whoami}, {\tt hostname} and {\tt netstat} are being used as stand-ins for
these data gathering applications. The collected data is written to
{\footnotesize \verb+C:\Users\User1\Documents\Thumbs\thumbit\test\thumbs.db+}.

\noindent {\em Data exfiltration}: Then the collected intelligence is
exfiltrated to 129.55.12.51:9418 using {\tt git}.

\noindent {\em Clean-up}: Dropper downloads a batch file called {\tt
  burnout.bat}. This file contains commands to clean up the attack footprint,
which are executed by {\tt cmd.exe} (see edges 11,12, 31-33).
\paragraph{Campaign F-3.} (Figure \ref{fig:pandex}).
Under the command of an attacker who uses stolen ssh
credentials, {\small \tt sshd} forks a {\small \tt bash} process. Note that
though there is no direct evidence from the audit data about the stolen ssh
credentials, because of the subsequent events ({\small \tt scp}) from this
shell, we conclude this as a sign of an attacker that uses stolen ssh
credentials.

Next the attacker,  invokes {\small \tt scp}, which downloads a file into location {\small \tt
  /var/dropbear\_latest/dropbearFREEBSD.tar}, which is then uncompressed. The
file {\small \tt dropbearscript} is next read and interpreted by {\small \tt
  sh}. This action creates the process {\small \tt dropbearkey}, which writes to
{\small \tt /usr/} {\small \tt local/etc/dropbear/dropbear\_ecdsa\_host\_key}
and {\small \tt /usr/local/etc/dropbear/dropbear\_rsa\_host\_key}. Next, another {\small \tt sudo} process
created by {\small \tt bash} starts another {\small \tt dropbear} process which
reads these two keys for future use (presumably to assist in connecting to a
remote host).

Dropbear next starts a shell process,
which executes a series of commands {\small \tt ls}, {\small \tt bash}, {\small
  \tt uname}, {\small \tt ps}, all of which write to a file {\small \tt
  /usr/home/user/procstats}.

 Finally, {\small \tt dropbear} starts a
          {\small \tt bash} process, which uses {\small \tt scp} to download a
          file called {\small \tt /usr/home/user/archiver}, and executes that
          file. The resulting process, called {\small \tt archiver}, reads the
          file {\small \tt /usr/home/user/procstats}, which contains the data
          output earlier, and exfiltrates this information to {\small \tt
            128.55.12.167:2525}.

\paragraph{Summary.} The above two graphs were constructed automatically by
\projname from audit data. They demonstrate how \projname enables an analyst to
obtain compact yet complete attack scenarios from hours of audit data. \projname
is able to hone in on the attack activity, even when it is hidden among benign
data that is at least three orders of magnitude larger.

\subsection{Overall Effectiveness}\label{subsec:effectiveness}
 \begin{table}[b]
 \begin{scriptsize}
 \begin{center}
 \scalebox{0.935}{
   \begin{tabular}{||c|c|c|c|c|c|c||}
      \hline
      \pbox{0.6cm}{Dataset} & \begin{tabular}[c]{@{}c}\pbox{0.2cm}{Drop \& \\ Load} \end{tabular} & \begin{tabular}[c]{@{}c} \pbox{0.8cm}{Intelligence \\Gathering} \end{tabular} & \begin{tabular}[c]{@{}c}\pbox{0.58cm}{Backdoor \\ Insertion}\end{tabular}  & \begin{tabular}[c]{@{}c}\pbox{0.65cm}{Privilege \\Escalation}\end{tabular}   & \begin{tabular}[c]{@{}c}\pbox{0.73cm}{Data \\Exfiltration} \end{tabular} & \begin{tabular}[c]{@{}c}\pbox{0.45cm}{Cleanup} \end{tabular}\\ [2.5ex]
      \hline \hline
      \ts W-1   &\checkmark &\checkmark &     &     &\checkmark &\checkmark\\ \hline
      \ts W-2   &\checkmark &\checkmark &\checkmark &     &\checkmark &\checkmark\\ \hline
       \ts L-1      &\checkmark &\checkmark &\checkmark &     &\checkmark &\checkmark\\ \hline
      \ts L-2     &\checkmark &\checkmark &\checkmark &\checkmark &\checkmark &\checkmark\\ \hline
      \ts L-3     &\checkmark &\checkmark &\checkmark &\checkmark &\checkmark &\checkmark\\ \hline
       \ts F-1  &     &     &\checkmark &     &\checkmark &\\ \hline
      \ts F-2     &\checkmark &\checkmark &\checkmark &     &\checkmark &\\ \hline
      \ts F-3   &\checkmark &\checkmark &     &     &\checkmark &\\ \hline   
   \end{tabular}}
 \end{center}
 \end{scriptsize}

 \caption{\projname results with respect to a typical APT campaign.}\label{apt-checklist}
  \end{table}

To assess the effectiveness of \projname in capturing essential stages of an
APT, in Table \ref{apt-checklist}, we correlate pieces of attack scenarios
constructed by \projname with APT stages documented in postmortem reports of
notable APT campaigns (e.g., the MANDIANT \cite{mandiant-report} report).
In 7 of the 8 attack scenarios, \projname uncovered the drop\&load activity. In
all the scenarios, \projname captured concrete evidence of data exfiltration, a
key stage in an APT campaign. In 7 of the scenarios, commands used by the
attacker to gather information about the target host were captured by \projname.

Another distinctive aspect of an APT is the injection of backdoors to targets
and their use for C\&C and data exfiltration. In this regard, 6 of the 8
scenarios reconstructed by \projname involve backdoor injection. Cleaning the
attack footprint is a common element of an APT campaign. In our experiments, in
5 of the 8 scenarios, \projname uncovered attack cleanup activities, e.g.,
removing dropped executables and data files created during the attack.

\begin{table}[t]
  \begin{scriptsize}
    \begin{center}
    \scalebox{0.935}{
      \begin{tabular}{||c|c|c|c|c|c|c|c||}
        \hline
         \begin{tabular}[c]{@{}c} \pbox[c]{0.35cm}{Dataset} \end{tabular} & \begin{tabular}[c]{@{}c} \pbox{0.4cm}{Entry\\ Entities}   \end{tabular} & \begin{tabular}[c]{@{}c} \pbox{0.55cm}{Programs \\Executed} \end{tabular} & \begin{tabular}[c]{@{}c}\pbox{0.1cm}{Key Files}\end{tabular}  & \begin{tabular}[c]{@{}c}\pbox{0.22cm}{Exit\\ Points}\end{tabular} & \begin{tabular}[c]{@{}c}\pbox{0.6cm}{Correctly\\Identified\\ Entities} \end{tabular} & \begin{tabular}[c]{@{}c}\pbox{0.7cm}{Incorrectly \\Identified \\Entities}\end{tabular}& \begin{tabular}[c]{@{}c@{}}\pbox{0.55cm}{Missed \\Entities}\end{tabular}\\ [2.5ex]
        \hline \hline
         \ts W-1            & 2  & 8  & 7  & 3  & 20 &0 & 0 \\ \hline 
        \ts  W-2            & 2  & 8  & 4  & 4  &18 & 0 & 0 \\ \hline \hline
        \ts L-1             & 2  & 10 &  7 & 2  & 20& 0 & 1\\ \hline 
         \ts  L-2           & 2  & 20 & 11 & 4 & 37& 0 & 0 \\ \hline
        \ts  L-3            & 1  & 6  &  6 & 5   &18 & 0 & 0\\ \hline \hline
        \ts F-1       & 4  &  13 & 9  & 2  & 13 &0 & 1\\ \hline 
        \ts F-2       & 2  & 10 & 7  & 3  & 22 & 0 & 0\\ \hline
        \ts F-3       & 4  & 14 & 7  & 1  &26 & 0 & 0\\ \hline \hline
        \ts{\bf Total} & {\bf 19} & {\bf 89 }& {\bf 58} & {\bf 24 }& {\bf174} & {\bf 0} &{\bf 2}\\ \hline
      \end{tabular}}
    \end{center}
  \end{scriptsize}
  \caption{Attack scenario reconstruction summary.}

\label{detection-summary}
\end{table}

Table \ref{detection-summary} shows another way of breaking down the attack
scenario reconstruction results, counting the number of key files, network
connections, and programs involved in the attack. Specifically, we count the
number of attack entry entities (including the entry points and the processes
that communicate with those entry points), attack-related program executions,
key files that were generated and used during the campaign, and the number of
exit points used for exfiltration (e.g., network sockets). This data was
compared with the ground truth, which was made available to us after we obtained
the results. The last two columns show the incorrectly reported and missed
entities, respectively. 

The two missed entities were the result of the fact that we had not spent any
effort in cataloging sensitive data files and device files. As a result, these
entities were filtered out during the forward analysis and simplification
steps. Once we marked the two files correctly, they were no longer filtered
out, and we were able to identify all of the key entities.

In addition to the missed entities shown in Table \ref{detection-summary}, the
red team reported that we missed a few other attacks and entities. 
Some of these were in data sets we did not examine. In particular, campaign W-2 was run multiple times, and we examined the data set from only one instance of it.  Also, there was a third attack campaign W-3 on Windows, but the team producing Windows data sets had difficulties during W-3 that caused the attack activities not to be recorded, so that data set is omitted from the results in Table \ref{detection-summary}.
Similarly, the team responsible for producing Linux data sets had 
some issues during campaign L-3 that caused some attack activities not to be recorded.  To account for this,
Table \ref{detection-summary} counts only the subset of key entities whose
names are present in the L-3 data set given to us.

According to the ground truth provided by the red team, we incorrectly
identified 21 entities in F-1 that were not part of an attack. Subsequent
investigation showed that the auditing system had not been shutdown at the end
of the F-1 campaign, and all of these false positives correspond
to testing/administration steps carried out after the end of the engagement, when the auditing system should not have been running.

\begin{table}[t!]
\begin{scriptsize}
\begin{center}
\begin{tabular}{||c|r|r|r|r|r||}
\hline
Dataset & \multicolumn{1}{l|}{\begin{tabular}[c]{@{}l@{}}Log Size\\on Disk\end{tabular}}  & \multicolumn{1}{l|}{\begin{tabular}[c]{@{}l@{}}\# of\\Events\end{tabular}} & \multicolumn{1}{l|}{\begin{tabular}[c]{@{}l@{}}Duration\\hh:mm:ss\end{tabular}}  & \multicolumn{1}{l|}{\begin{tabular}[c]{@{}l@{}}Packages\\Updated\end{tabular}} & \multicolumn{1}{l||}{\begin{tabular}[c]{@{}l@{}}Binary\\Files\\Written\end{tabular}} \\ \hline
\begin{tabular}[c]{@{}l@{}}Server 1\end{tabular} &1.1G   & 2.17M  & 00:13:06     & 110     & 1.8K  \\ \hline
\begin{tabular}[c]{@{}l@{}}Server 2\end{tabular} &2.7G   & 4.67M  & 105:08:22    & 4       & 4.2K  \\ \hline
\begin{tabular}[c]{@{}l@{}}Server 3\end{tabular} &12G    & 20.9M  & 104:36:43    & 4       & 4.3K  \\ \hline
\begin{tabular}[c]{@{}l@{}}Server 4\end{tabular} &3.2G   & 5.09M  & 119:13:29    & 4       & 4.3K       \\ \hline
\end{tabular}\caption{False alarms in a benign environment with software upgrades and updates. No alerts were triggered during this period.} \label{benignTabl}
\end{center}
\end{scriptsize}
\end{table}

\subsection{False Alarms in a Benign Environment}
\label{fabe}
In order to study \projname's performance in a benign environment, we collected
audit data from four Ubuntu Linux servers over a period of 3 to 5 days. One of
these is a mail server, another is a web server, and a third is an NFS/SSH/SVN
server. Our focus was on software updates and upgrades during this period, since
these updates can download code from the network, thereby raising the
possibility of untrusted code execution alarms. There were four security updates
(including kernel updates) performed over this period. In addition, on a fourth
server, we collected data when a software upgrade was performed, resulting in
changes to 110 packages. Several thousand binary and script files were updated
during this period, and the audit logs contained over 30M events. All of this
information is summarized in Table~\ref{benignTabl}.

As noted before, policies should be configured to permit software updates and
upgrades using standard means approved in an enterprise. For Ubuntu Linux, we
had one policy rule for this: when \mytt{dpkg} was executed by
\mytt{apt}-commands, or by \mytt{unattended-upgrades}, the process is not
downgraded even when reading from files with untrusted labels. This is because
both \mytt{apt} and \mytt{unattended-upgrades} verify and authenticate the hash
on the downloaded packages, and only after these verifications do they invoke
\mytt{dpkg} to extract the contents and write to various directories containing
binaries and libraries. Because of this policy, all of the 10K+ files downloaded
were marked benign. As a result of this, no alarms were generated from their
execution by \projname.
\subsection{Runtime and Memory Use}

\begin{table}
  \begin{center}
\begin{adjustbox}{width=0.95\columnwidth}
 \begin{scriptsize}
  \begin{tabular}{||c|c|r|c|r||}
    \hline
     \ts Dataset & \ts Duration & \ts Memory & \multicolumn{2}{|c||}{\ts Runtime}  \\
    \cline{4-5} & (hh:mm:ss) &Usage& \ts Time & Speed-up \\
    \hline \hline
     \ts W-1 & 06:22:42&3 MB & 1.19 s & 19.3 K  \\ \hline 
    \ts W-2 & 19:43:46 &10 MB  & 2.13 s & 33.3 K  \\ \hline \hline
    \multicolumn{2}{||c}{\ts \bf W-Mean}& 6.5 MB &\multicolumn{2}{r|}{\ts \bf 26.3 K}  \\ \hline\hline
    \ts L-1  & 07:59:26&26 MB & 8.71 s  & 3.3 K  \\ \hline 
    \ts L-2  & 79:06:39&329 MB & 114.14s & 2.5 K  \\ \hline
    \ts L-3  & 79:05:13&175 MB & 74.14 s & 3.9 K   \\ \hline \hline
    \multicolumn{2}{||c}{\ts \bf L-Mean} &\ts 177 MB &\multicolumn{2}{r|}{\ts \bf 3.2 K} \\ \hline
     \ts F-1 & 08:17:30&8 MB & 1.86 s & 16 K   \\ \hline 
     \ts F-2 & 78:56:48&84 MB & 14.02 s & 20.2 K  \\ \hline
     \ts F-3 & 79:04:54&95 MB & 15.75 s & 18.1 K \\ \hline \hline
    \multicolumn{2}{||c}{\ts \bf F-Mean}&62.3 MB &\multicolumn{2}{r|}{\ts \bf 18.1 K} \\ \hline
  \end{tabular}
  \end{scriptsize}
  \end{adjustbox}
  \end{center}
 \caption{Memory use and runtime for scenario reconstruction. } \label{tab:perf}
\end{table}

\rtodo{Add the column about the scenario reconstruction in table 6: Memory use
  and runtime for scenario reconstruction.} Table~\ref{tab:perf} shows the
runtime and memory used by \projname for analyzing various scenarios. The
measurements were made on a Ubuntu 16.04 server with 2.8GHz AMD Opteron 62xx
processor and 48GB main memory. Only a single core of a single processor was
used. The first column shows the campaign name, while the second shows the total
duration of the data set.

The third column shows the memory used for the dependence graph. As described in
Section~\ref{sec:mainmem}, we have designed a main memory representation that is
very compact. This compact representation enables \projname to store data
spanning very long periods of time. As an example, consider campaign L-2, whose
data were the most dense. \projname used approximately 329MB to store 38.5M
events spanning about 3.5 days. Across all data sets, \projname needed about 8
bytes of memory per event on the larger data sets, and about 20 bytes per event
on the smaller data sets.

 \begin{table*}[ht]
  \begin{footnotesize}
  \begin{center}
  \begin{tabular}{||c|c|c|c|c|c|c|c|c||}
  \hline
  \multirow{2}{*}{Dataset} & \multicolumn{2}{c|}{\begin{tabular}[c]{@{}c@{}}Untrusted\\ execution\end{tabular}} & \multicolumn{2}{c|}{\begin{tabular}[c]{@{}c@{}}Modification by\\ low code t-tag subject\end{tabular}} & \multicolumn{2}{c|}{\begin{tabular}[c]{@{}c@{}}Preparation of untrusted\\ data for execution\end{tabular}} & \multicolumn{2}{c||}{\begin{tabular}[c]{@{}c@{}}Confidential \\ data leak\end{tabular}} \\ \cline{2-9} 
       & \multicolumn{1}{c|}{Single t-tag}   & \multicolumn{1}{c|}{Split t-tags} & \multicolumn{1}{l|}{Single t-tag}  & \multicolumn{1}{c|}{Split t-tags} & \multicolumn{1}{c|}{Single t-tags} & \multicolumn{1}{c|}{Split t-tags} & \multicolumn{1}{c|}{Single t-tag} & \multicolumn{1}{c||}{Split t-tags}  \\ \hline
    \ts  W-1   & 21    & 3  & 1.2 K     & 3     & 0    & 0   & 6.1 K    & 11  \\ \hline 
    \ts  W-2   & 44    & 2  & 3.7 K   & 108   & 0    & 0   & 20.2 K   & 18  \\ \hline \hline
     \ts L-1   & 60    & 2  & 53      & 5     & 1    & 1   & 19       & 6     \\ \hline
    \ts L-2    & 1.5 K & 5  & 19.5 K  & 1     & 280  & 8   & 122 K    & 159   \\ \hline
    \ts L-3    & 695   & 5  & 26.1 K  & 2     & 270  & 0   & 62.1 K   & 5.3 K \\ \hline \hline
\multicolumn{2}{||c}{\ts \bf  Average Reduction}& {\ts \bf 45.39x} &\multicolumn{2}{r|}{\ts \bf 517x} &\multicolumn{2}{r|}{\ts \bf 6.24x} &\multicolumn{2}{r||}{\ts \bf 112x}\\ \hline
  \end{tabular}
  \caption{Reduction in (false) alarms by maintaining separate code and data
trustworthiness tags. The average reduction shows the average factor of reduction we get for alarms generation when using split trustworthiness tag over single trustworthiness tag.} \label{alarm}
  \end{center}
  \end{footnotesize}
  \end{table*}

The fourth column shows the total run time, including the times for consuming
the dataset, constructing the dependence graph, detecting attacks, and
reconstructing the scenario. We note that this time was measured after the
engagement when all the data sets were available. During the engagement,
\projname was consuming these data as they were being produced. Although the
data typically covers a duration of several hours to a few days, the analysis
itself is very fast, taking just seconds to a couple of minutes. Because of our
use of tags, most information needed for the analysis is locally available. This
is the principal reason for the performance we achieve. 

The ``speed-up'' column illustrates the performance benefits of \projname. It
can be thought of as the number of simultaneous data streams that can be handled
by \projname, if CPU use was the only constraint.

{\em In summary}, \projname is able to consume and analyze audit COTS data from
several OSes in real time while having a small memory footprint.

\begin{table}[b]
\begin{scriptsize}
\begin{center}
\begin{tabular}{||c|c|c|c|c|c|c||}
\hline
\multirow{2}{*}{\ts \pbox{0.9cm}{Dataset}} & \multirow{2}{*}{\begin{tabular}[c]{@{}c@{}}Initial\\ \# of \\Events\end{tabular}}  & \multirow{2}{*}{\begin{tabular}[c]{@{}c@{}}Final\\ \# of\\ Events\end{tabular}}& \multicolumn{4}{c||}{{Reduction Factor}} \\ \cline{4-7}
& &  & \begin{tabular}[c]{@{}c@{}}Single\\ t-tag\end{tabular} & \begin{tabular}[c]{@{}c@{}}Split\\ t-tag\end{tabular} & \begin{tabular}[c||]{@{}c@{}}\projname\\ Simplif.\end{tabular} & Total     \\ \hline \hline

\ts W-1       &  100 K & 51  & 4.4x  & 1394x  & 1.4x &   1951x \\ \hline 
\ts W-2       &  401 K & 28  & 3.6x  & 552x   & 26x  &  14352x \\ \hline \hline
\ts L-1       & 2.68 M & 36  & 8.9x  & 15931x & 4.7x &  74875x\\ \hline 
\ts L-2       & 38.5 M &130  & 7.3x  & 2971x  & 100x & 297100x\\ \hline
\ts L-3       & 19.3 M & 45  & 7.6x  & 1208x  & 356x & 430048x\\ \hline \hline
\ts F-1       &  701 K & 45  & 2.3x  & 376x   & 41x  &  15416x\\ \hline %
\ts F-2       & 5.86 M & 39  & 1.9x  & 689x   & 218x & 150202x\\ \hline%
\ts F-3       & 5.68 M & 45  & 6.7x  & 740x   & 170x & 125800x\\ \hline  \hline %
\multicolumn{3}{||c}{\ts \bf Average Reduction}& \multicolumn{1}{r|}{\ts \bf 4.68x} &\multicolumn{1}{c|}{\ts \bf 1305x} &\multicolumn{1}{c|}{\ts \bf 41.8x} &\multicolumn{1}{r||}{\ts \bf 54517x} \\ \hline 

\end{tabular}
\caption{Comparison of selectivity achieved using forward analysis with single trustworthiness tags, forward analysis with split code and data trustworthiness tags, and finally simplifications. \label{selectivity-all}}
\end{center}
\end{scriptsize}
\end{table}
\rtodo{Changed the numbers in the last column in the table 8 (and the factors) to make sure that they match the graphs. There is one oddity, 3 attacks have the same number of events, 45. This might be seen as unlikely. Old table is commented out in latex.}

\subsection{Benefit of split tags for code and data}
\label{Ss:codedatatags}
As described earlier, we maintain two trustworthiness tags for each subject, one
corresponding to its code, and another corresponding to its data. By
prioritizing detection and forward analysis on code trustworthiness, we
cut down vast numbers of alarms, while greatly decreasing the size of 
forward analysis output. 

Table~\ref{alarm} shows the difference between the number of alarms generated
by our four detection policies with single trustworthiness tag and with the split trustworthiness (code and integrity) tags.
Note that the split reduces the alarms by a factor of 100 to over 1000 in
some cases. 

\begin{figure}[b]
\begin{center}
\includegraphics[scale=0.4]{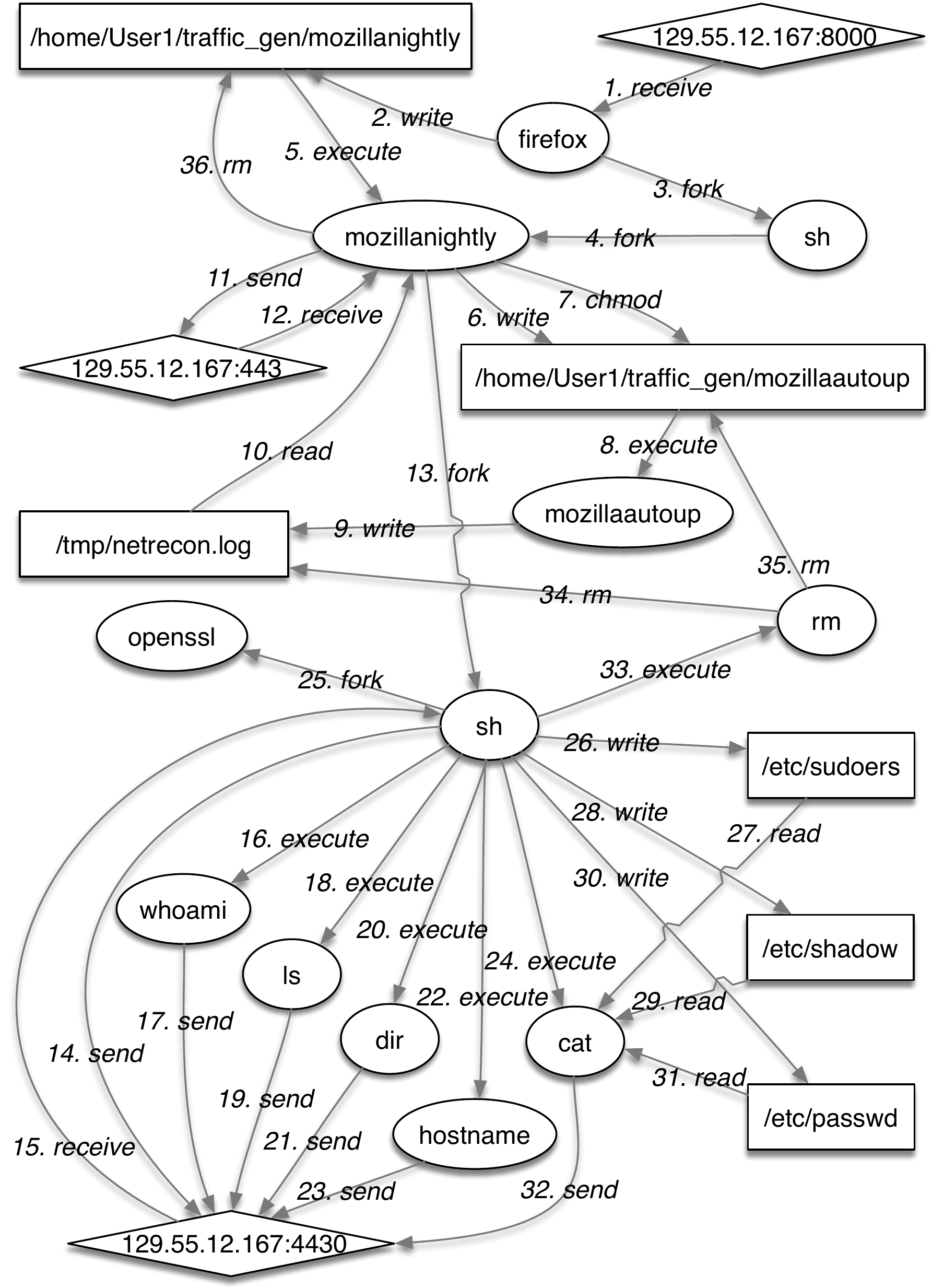}
\end{center}
\caption{Scenario graph reconstructed from campaign L-1.}
\label{sristr_2}
\end{figure}

\begin{figure*}
\begin{center}
\includegraphics[scale=0.4]{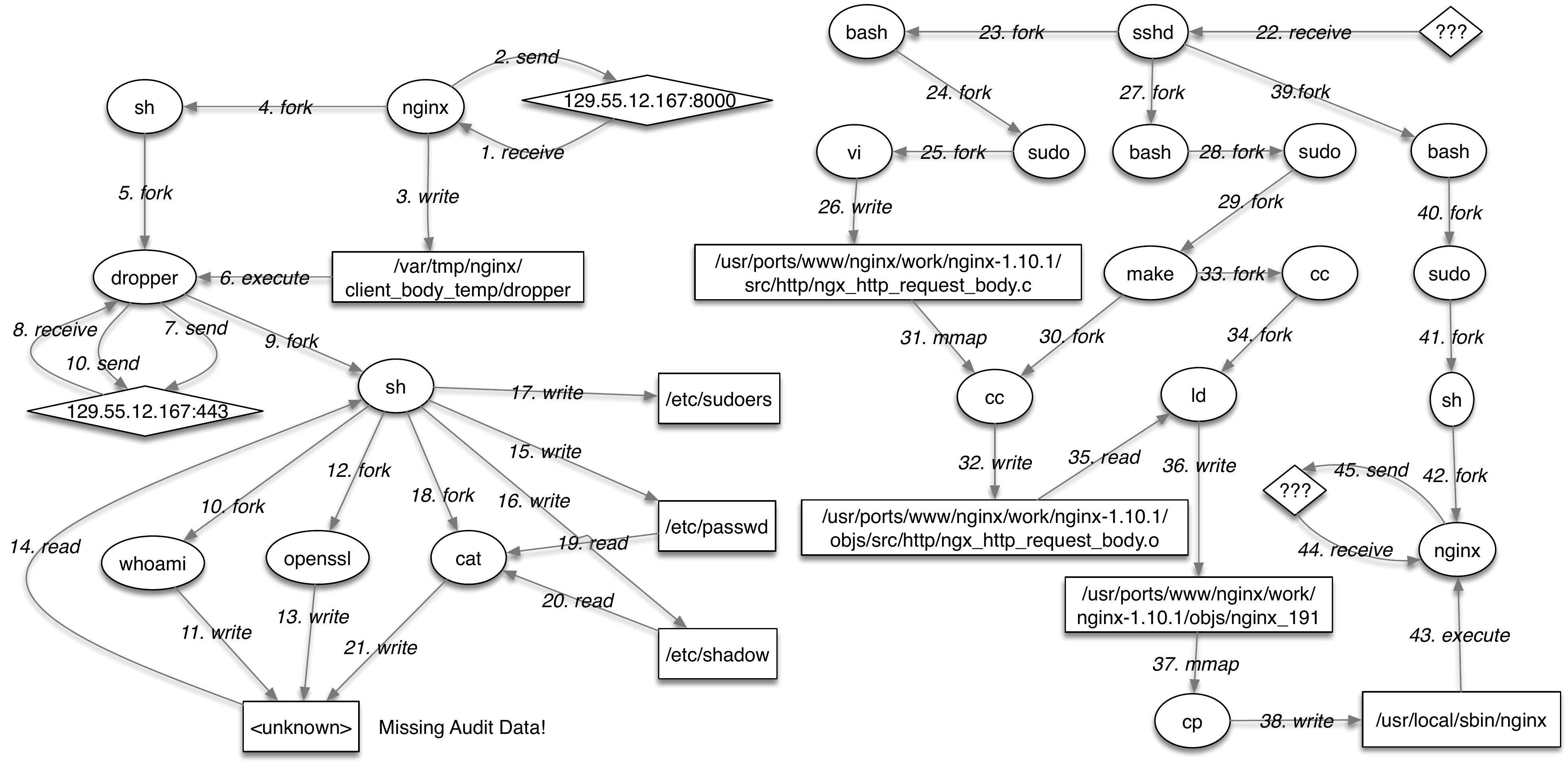}
\end{center}
\caption{ Scenario graph reconstructed from campaign F-1.}
\label{cadets-stretch}
\end{figure*}

Table~\ref{selectivity-all} shows the improvement achieved in forward analysis
as a result of this split. In particular, the increased selectivity reported
in column 5 of this table comes from splitting the tag. Note that often,
there is a 100x to 1000x reduction in the size of the graph.

\subsection{Analysis Selectivity}\label{sec:selectivity}

Table \ref{selectivity-all} shows the data reduction pipeline of the analyses in
\projname. The second column shows the number of original events in each
campaign. These events include all the events in the system (benign and
malicious) over several days with an overwhelming majority having a benign
nature, unrelated to the attack. 

The third column shows the final number of events that go into the attack
scenario graph.

The fourth column shows the reduction factor when a naive forward analysis with
single trustworthiness tag (single t-tag) is used from the entry points
identified by our backward analysis. Note that the graph size is very large in
most cases. The fifth column shows the reduction factor using the forward
analysis of \projname --- which is based on split (code and data) trustworthiness
tags. As can be seen from the table, \projname achieved two to three orders of
magnitude reduction with respect to single t-tag based analysis.

The output of forward analysis is then fed into the simplification engine. The
sixth column shows the reduction factor achieved by the simplifications over the
output of our forward analysis.
The last column shows the overall reduction we get over original events using
split (code and data) trustworthiness tags and performing the simplification.

Overall, the combined effect of all of these steps is very substantial: data
sets consisting of tens of millions of edges are reduced into graphs with
perhaps a hundred edges, representing five orders of magnitude reduction in the
case of L-2 and L-3 data sets, and four orders of magnitude reduction on other
data.

\subsection{Discussion of Additional Attacks}\label{subsec:additional}
In this section, we provide graphs that reconstruct attack
campaigns that weren't discussed in Section~\ref{sec:selres}.
Specifically, we discuss attacks L-1, F-1, F-2, W-1, and L-3.
\textbf{Attack L-1.} In this attack (Figure \ref{sristr_2}), \texttt{firefox} is exploited to drop and execute via a shell the file \texttt{mozillanightly}. The process \texttt{mozillanightly} first downloads and executes \texttt{mozillaautoup}, then starts a shell, which spawns several other processes. Next, the information gathered in file \texttt{netrecon.log} is exfiltrated and the file removed.

\textbf{Attack F-1.} In this attack (Figure \ref{cadets-stretch}), the \texttt{nginx} server is exploited to drop and execute via shell the file \texttt{dropper}. Upon execution, the \texttt{dropper} process forks a shell that spawns several processes, which write to a file and reads and writes to sensitive files. In addition, \texttt{dropper} communicates with the IP of the attacker. We report in the figure the graph related to the restoration and administration carried out after the engagement, as discussed in Section \ref{subsec:effectiveness}.

\begin{figure*}[h]
\begin{center}
\vspace*{-2em}
\includegraphics[scale=0.35]{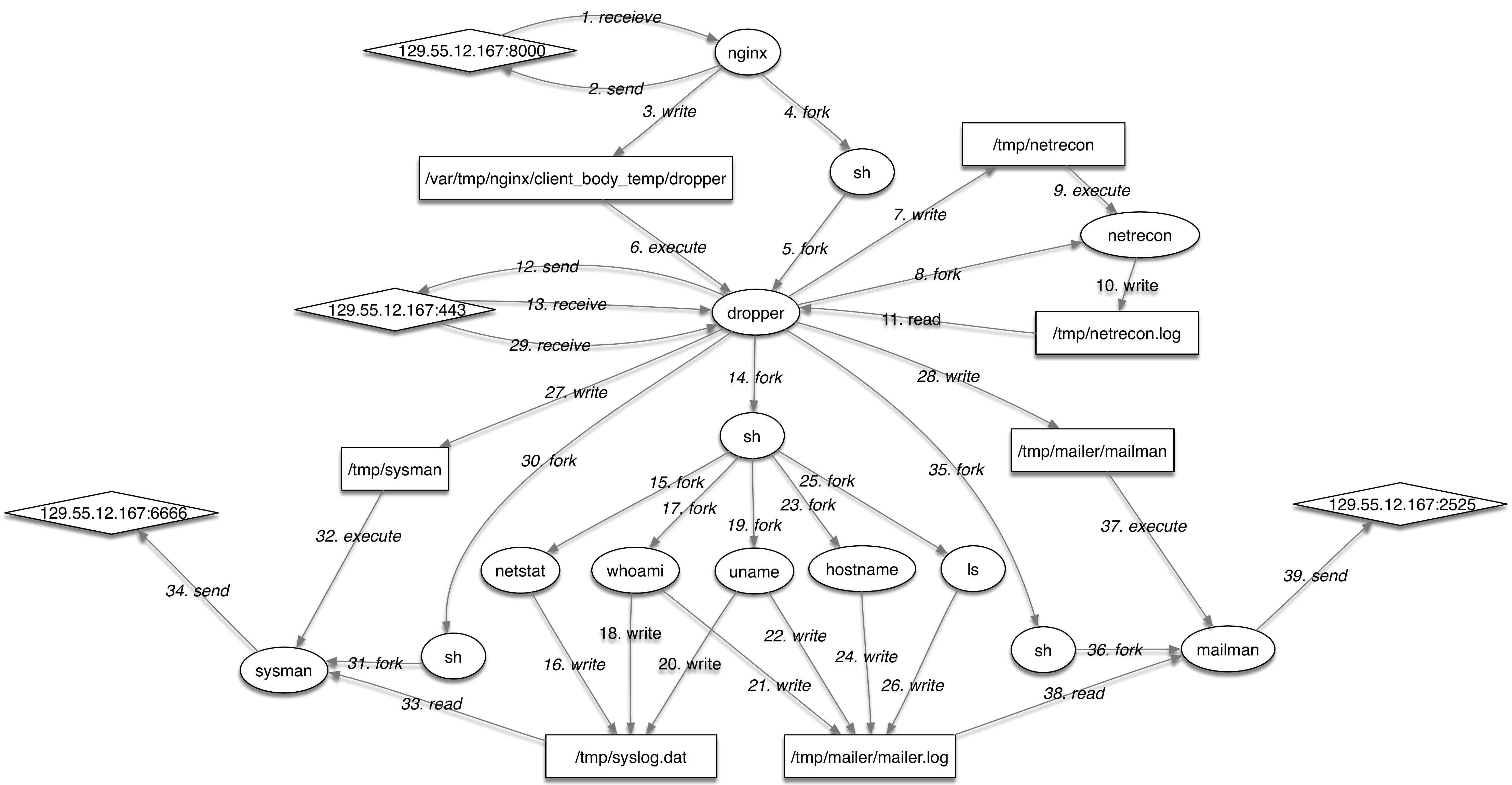}
\end{center}
\caption{ Scenario graph reconstructed from campaign F-2.}
\label{cadets-bovia}
\end{figure*}

\textbf{Attack F-2.} The start of this attack (Figure \ref{cadets-bovia}) is similar to F-1. However, upon execution, the \texttt{dropper} process downloads three files named \texttt{recon, sysman}, and \texttt{mailman}. Later, these files are executed and used which are used to exfiltrate data gathered from the system. 

\begin{figure*}
\begin{center}
\vspace*{1em}
\includegraphics[scale=0.36]{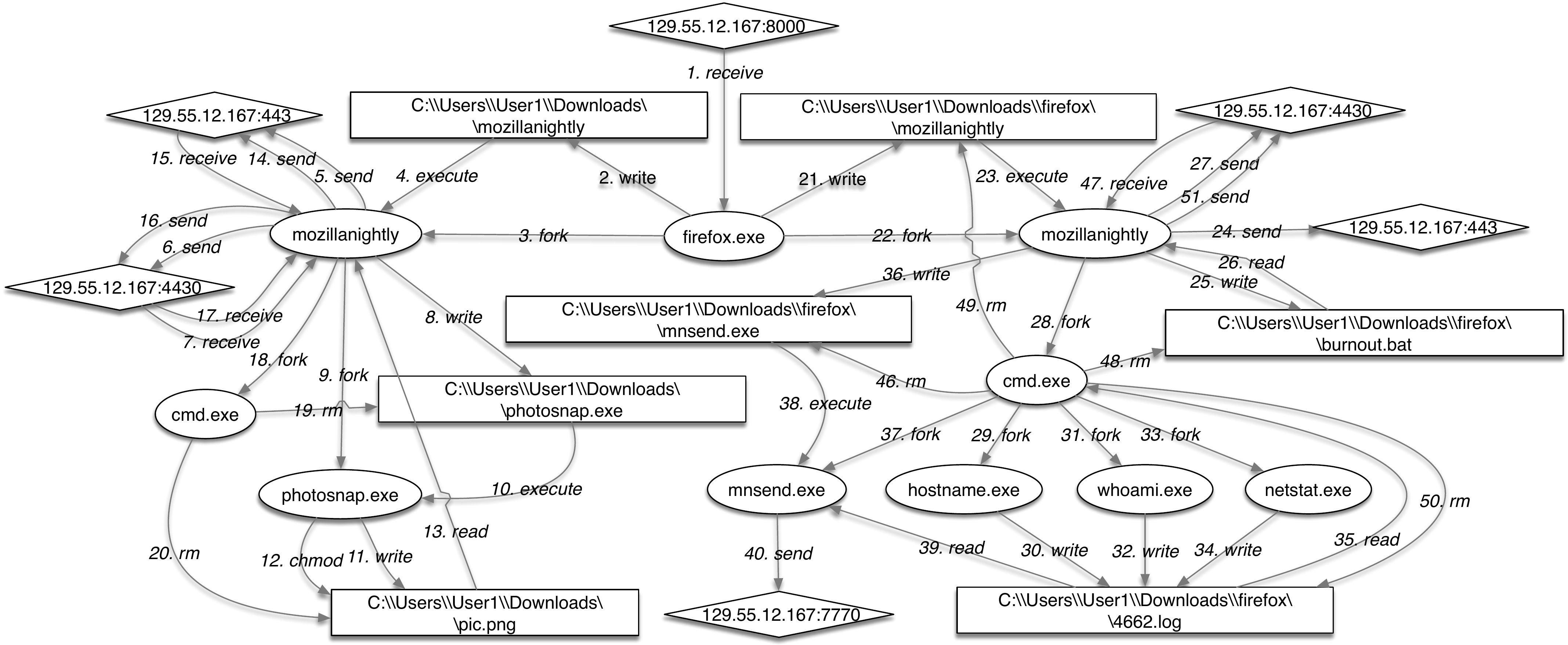}
\end{center}
\caption{ Scenario graph reconstructed from campaign W-1.}
\label{5d-stretch}
\end{figure*}

\begin{figure*}
\begin{center}
\includegraphics[scale=0.38]{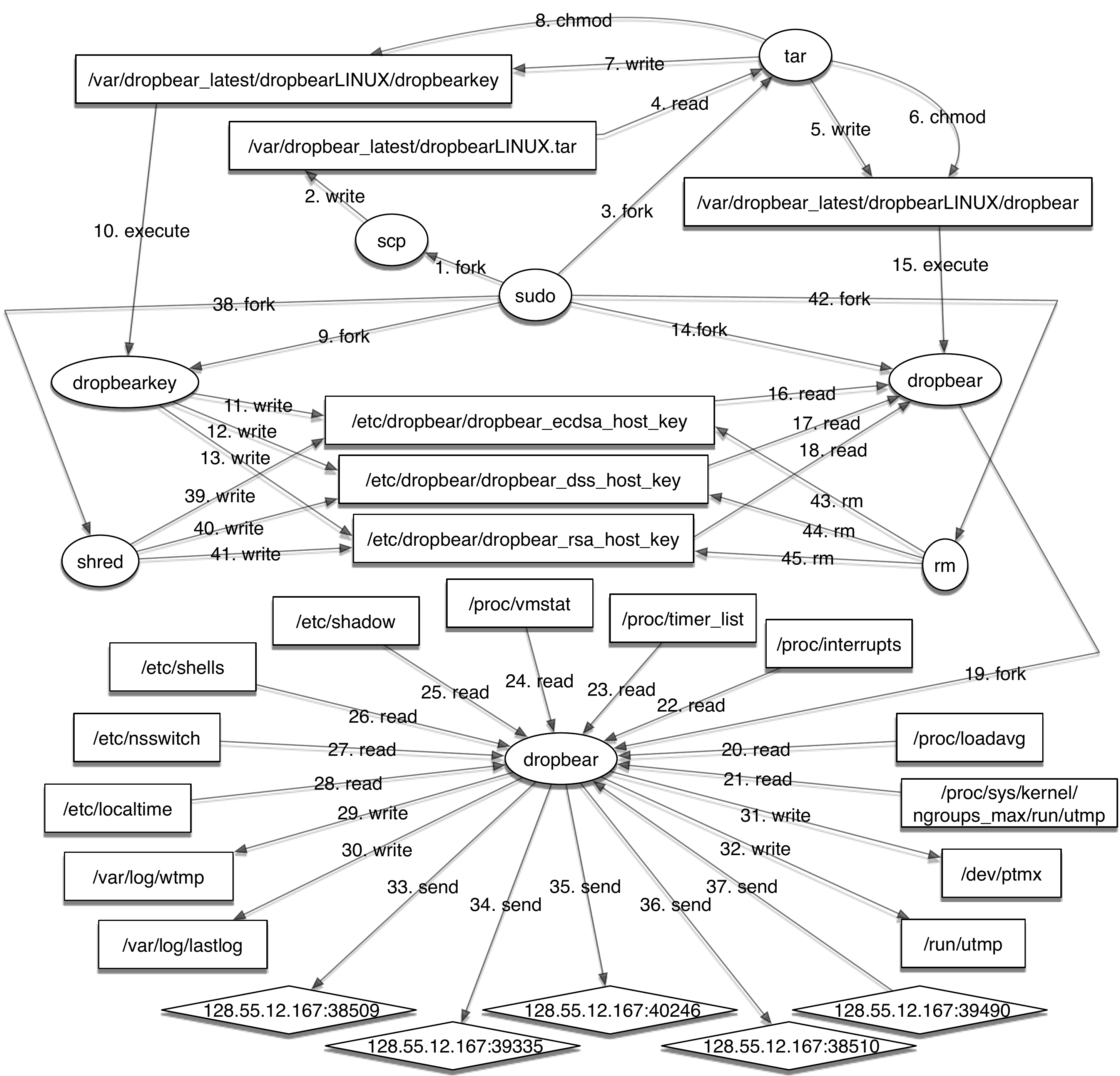}
\end{center}
\caption{ Scenario graph reconstructed from campaign L-3.}
\label{sri-pandex}
\end{figure*}

\textbf{Attack W-1}. In this attack (Figure \ref{5d-stretch}), \texttt{firefox} is exploited twice to drop and execute a file \texttt{mozillanightly}. The first \texttt{mozillanightly} process downloads and executes the file \texttt{photosnap.exe}, which takes a screenshot of the victim's screen and saves it to a png file. Subsequently, the jpeg file is exfiltrated by \texttt{mozillanightly}. The second \texttt{mozillanightly} process downloads and executes two files: 1) \texttt{burnout.bat}, which is read, and later used to issue commands to \texttt{cmd.exe} to gather data about the system; 2) \texttt{mnsend.exe}, which is executed by \texttt{cmd.exe} to exfiltrate the data gathered previously.

\textbf{Attack L-3.} In this attack (Figure \ref{sri-pandex}), the file \texttt{dropbearLINUX.tar} is downloaded and extracted. Next, the program \texttt{dropbearkey} is executed to create three keys, which are read by a program \texttt{dropbear}, which subsequently performs exfiltration.

\section{Related Work}
\label{sec:relw}
In this section, we compare \projname with efforts from academia and open source
industry tools. We omit comparison to proprietary products from the industry as
there is scarce technical documentation available for an in-depth comparison.

\paragraph{Provenance tracking and Forensics} Several logging and provenance tracking systems have been built to monitor the activities of a system 
\cite{gehani2012spade,muniswamy2006provenance,taser2005,forensix,Braun2006,
  pohly2012hi,bates2015trustworthy} and build \textit{provenance graphs}. Among
these, \textit{Backtracker} \cite{king2003backtracking,king2005enriching} is one
of the first works that used dependence graphs to trace back to the root causes
of intrusions. These graphs are built by correlating events collected by a
logging system and by determining the causality among system entities, to help
in forensic analysis after an attack is detected.

\projname improves on the techniques of Backtracker in two important ways.
First, Backtracker was meant to operate in a forensic setting, whereas our
analysis and data representation techniques are designed towards real-time
detection. Setting aside hardware comparisons, we note that Bactracker took 3
hours for analyzing audit data from a 24-hour period, whereas \projname was able
to process 358 hours of logs in a little less than 3 minutes. Secondly,
Backtracker relies on alarms generated by external tools, therefore its forensic
search and pruning cannot leverage the reasons that generated those alarms. In
contrast, our analysis procedures leverage the results from our principled
tag-based detection methods and therefore are inherently more precise. For
example, if an attack deliberately writes into a well-known log file,
Backtracker's search heuristics may remove the log file from the final graph,
whereas our tag-based analysis will prevent that node from being pruned away.

In a similar spirit, \textit{BEEP} \cite{lee2013high} and its
  evolution \textit{ProTracer} \cite{ma2016protracer} build dependence graphs
  that are used for forensic analysis. In contrast, SLEUTH builds dependence
  graphs for real-time detection from which scenario subgraphs are extracted
  during a forensic analysis. The forensic analysis of
  \cite{lee2013high,ma2016protracer} ensures more precision than
  Backtracker~\cite{king2003backtracking} by heuristically dividing the
  execution of the program into execution units, where each unit represents one
  iteration of the main loop in the program. The instrumentation required to
  produce units is not always automated, making the scalability of their
  approach a challenge. \projname can make use of the additional precision
  afforded by \cite{lee2013high} in real-time detection, when such information
  is available.

\comment{ for As a result, the forensic capability of our system provides much
  more precise description of the attack. In addition, to alleviate the
  dependency explosion problem, our system uses \textit{versioning} and tags. In
  contrast to \textit{BEEP} and \textit{ProTracer} our \textit{versioning}
  approach does not require program instrumentation and it can be used also on
  programs that do not have a main-loop structure. Furthermore,
  \textit{versioning} is used on objects as well, providing additional
  improvements to the problem of dependency explosion. }

While the majority of the aforementioned systems operate at the system call
level, several other systems track information flows at finer granularities
\cite{angelos,flowdroid, lee2013high}. They typically instrument applications
(e.g., using Pin \cite{PIN}) to track information flows through a program. Such
fine-grained tainting can provide much more precise provenance information, at
the cost of higher overhead. Our approach can take advantage of finer
granularity provenance, when available, to further improve accuracy.

\comment{
  Note: I don't believe versioning does any thing to the actual accuracy we
  obtain in the evaluation. So, we should not bring it up here. The reason
  for the use of versioning is rather different from units, so let us
  avoid the comparison.
  
\vtodo{I'm not sure versioning and units are directly comparable. we should not
  get into such comparisons. The positive way to write it is as given in the
  note above, that we can work with any fine grained data if available. }

\todo{Rigel: I think versioning is one of the major contributions of the paper
  to fight dependency explosion, so we need to compare it with BEEP, which is
  also about fighting dependency explosion. I removed the `comparable
  improvements' however}

\vtodo{there is also work on eidetic systems :
  https://www.usenix.org/conference/osdi14/technical-sessions/presentation/devecsery

  here the goal is full system replay, whereas our goal is security audit}
}
\paragraph{Attack Detection} A number of recent research efforts on attack detection/prevention focus on
``inline'' techniques that are incorporated into the protected system, e.g.,
address space randomization, control-flow integrity, taint-based defenses and so
on. Offline {\em intrusion detection} using logs has been studied for a much
longer period \cite{denning1987intrusion,lunt1992real, senseofself}. In
particular, {\em host-based IDS} using system-call monitoring and/or audit logs
has been investigated by numerous research efforts \cite{alternative,
  datamining, sp01, idsbysa, vtpath, kruegelbook}.

Host-based intrusion detection techniques mainly fall into three categories: (1)
{\em misuse-based}, which rely on specifications of bad behaviors associated
with known attacks; (2) {\em anomaly-based} \cite{senseofself, datamining, sp01,
  execgraph, kruegelarg, detectionWindows2015, xiaokuiccs}, which rely on
learning a model of benign behavior and detecting deviations from this behavior;
and (3) {\em specification-based} \cite{ko1997execution, raid01}, which rely on
specifications (or policies) specified by an expert. The main drawback of
misuse-based techniques is that their signature-based approach is not amenable
to detection of previously unseen attacks. Anomaly detection techniques avoid
this drawback, but their false positives rates deter widespread deployment.
Specification/policy-based techniques can reduce these false positives, but they
require application-specific policies that are time-consuming to develop and/or
rely on expert knowledge. Unlike these approaches, \projname relies on {\em
  application-independent policies.} We develop such policies by exploiting
provenance information computed from audit data. In particular, an audit event

\paragraph{Information Flow Control (IFC)}  
IFC techniques assign security labels and propagate them in a manner similar to
our tags. %
Early works, such as Bell-LaPadula \cite{Bell:LaPadula} and Biba
\cite{biba77integrity}, relied on strict policies. These strict policies impact
usability and hence have not found favor among contemporary OSes. Although IFC
is available in SELinux \cite{selinux}, it is not often used, as users prefer
its access control framework based on domain-and-type enforcement. While most
above works centralize IFC, {\em decentralized IFC} (DIFC) techniques
\cite{histar, asbestos, flume} emphasize the ability of principals to define and
create new labels. This flexibility comes with the cost of nontrivial changes to
application and/or OS code. %

Although our tags are conceptually similar to those in IFC systems, the central
research challenges faced in these systems are very different from \projname. In
particular, the focus of IFC systems is enforcement and prevention.
A challenge for IFC enforcement is that their policies tend to break
applications. Thus, most recent efforts \cite{ppi, ifedac, Li:Mao:UMIP, spif,sze2014towards,sze2013portable,sun2008expanding} in
this regard focus on refinement and relaxation of policies so that compatibility
can be preserved without weakening security. In contrast, neither enforcement
nor compatibility pose challenges in our setting. On the other hand, IFC systems
do not need to address the question of what happens when policies are violated.
Yet, this is the central challenge we face: how to distinguish attacks from the
vast number of normal activities on the system; and more importantly, once
attacks do take place, how to tease apart attack actions from the vast amounts
of audit data.

\paragraph{Alert Correlation} 
Network IDSs often produce myriad alerts. {\em Alert correlation} analyzes
relationships among alerts, to help users deal with the deluge. The main
approaches, often used together, are to {\em cluster} similar alerts, prioritize
alerts, and identify causal relationships between alerts
\cite{debar2001,ning2003,qin2003,noel2004,wang2008}. Furthermore, they require
manually supplied expert knowledge about dependencies between alert types (e.g.,
consequences for each network IDS alert type) to identify causal relationships.
In contrast, we are not interested in clustering/statistical techniques to
aggregate alerts. Instead, our goals are to use provenance tracking to determine
causal relationships between different alarms to reconstruct the attack
scenario, and to do so without relying on (application-dependent) expert
knowledge.
\section{Conclusion} \label{conclu}
We presented an approach and a system called \projname for real-time
detection of attacks and attack reconstruction from COTS audit logs. \projname
uses a main memory graph data model and a rich tag-based policy framework that
make its analysis both efficient and precise. We evaluated \projname on large
datasets from 3 major OSes under attack by an independent red team, efficiently
reconstructing all the attacks with very few errors.

\begin{footnotesize}
\setlist[1]{leftmargin=1.2\parindent, itemsep=0ex, topsep=0ex, parsep=0ex%
}
\bibliographystyle{plain}
\bibliography{main}
\end{footnotesize}

\end{document}